\documentclass[12pt]{iopart}
\usepackage{amssymb}
\usepackage{epsfig,newlfont}
\usepackage{subfigure}
\usepackage[colorlinks, 
linkcolor=blue,
urlcolor=blue,
citecolor=blue,
bookmarksopen=false]{hyperref}
\usepackage{graphicx, xcolor}
\usepackage{caption}
\newcommand{\fsa}[1]{\left\langle{#1}\right\rangle}

\begin{document}

\title[Fast neoclassical simulations for large aspect ratio stellarators with KNOSOS]{Fast simulations for large aspect ratio stellarators with the neoclassical code KNOSOS}

\author{J L Velasco$^1$, I Calvo$^1$, F I Parra$^2$, V d'Herbemont$^{2,3}$, H M Smith$^4$, D Carralero$^1$, T Estrada$^1$ and the W7-X team}

\address{$^1$ Laboratorio Nacional de Fusi\'on, CIEMAT, Madrid, Spain}

\address{$^2$ Rudolf Peierls Centre for Theoretical Physics, University of Oxford, Oxford, United Kingdom}

\address{$^3$ Mines ParisTech, Paris, France}

\address{$^4$ Max-Planck Institut f\"ur Plasmaphysik, Greifswald, Germany}

\ead{joseluis.velasco@ciemat.es}

\vspace{10pt}
\begin{indented}
\item[]\today
\end{indented}

\begin{abstract}

In this work, a new version of \texttt{KNOSOS} is presented. \texttt{KNOSOS} is a low-collisionality radially-local, bounce-averaged neoclassical code that is extremely fast, and at the same time, includes physical effects often neglected by more standard codes: the component of the magnetic drift that is tangent to the flux-surface and the variation of the electrostatic potential on the flux-surface. An earlier version of the code could only describe configurations that were sufficiently optimized with respect to neoclassical transport. \texttt{KNOSOS} can now be applied to any large aspect ratio stellarator, and its performance is demonstrated by means of detailed simulations in the configuration space of Wendelstein 7-X.

\end{abstract}

\section{Introduction}

An accurate calculation of radial neoclassical transport is important for both tokamaks and stellarators. In tokamaks, deviations of the magnetic field from axisymmetry (caused, for example, by ripple due to the finite number of coils or by resonant magnetic perturbations) can result in significant neoclassical damping of the toroidal rotation~\cite{shaing2015overview}. In stellarators, their intrinsically three-dimensional configurations leads to specific neoclassical transport regimes (see e.g.~\cite{beidler2011ICNTS,calvo2017sqrtnu}) that produce radial energy transport comparable, and often larger, than its turbulent counterpart~\cite{dinklage2013ncval}. Although typically less demanding than gyrokinetic codes, the computational cost of neoclassical simulations is crucial for a thorough characterization of transport in three-dimensional configurations, especially at low plasma collisionalities. In this work we present a new version of \texttt{KNOSOS}~\cite{velasco2020knosos} (KiNetic Orbit-averaging SOlver for Stellarators), a freely-available open-source code that provides a fast computation of low collisionality neoclassical transport in three-dimensional magnetic confinement devices by rigorously solving the radially local bounce-averaged drift kinetic equation coupled to the quasineutrality equation. Apart from its remarkable speed, \texttt{KNOSOS}~\cite{velasco2020knosos} includes physics often neglected in neoclassical codes, such as the effect of the component of the magnetic drift that is tangent to magnetic surfaces and the component of the electrostatic potential that varies on the magnetic surface, $\varphi_1$~\cite{regana2018phi1} (only recently did stellarator neoclassical codes start to calculate $\varphi_1$, and at a large computational cost). The latter quantity can have a strong impact on the radial transport of highly-charged impurities in three-dimensional magnetic configurations~\cite{calvo2019nf}.  The fast calculation of the bounce averaged main ion distribution with \texttt{KNOSOS} opens the door to a fast evaluation of neoclassical impurity transport using recently-derived analytical expressions~\cite{calvo2018nf,calvo2019nf}.

An earlier version of \texttt{KNOSOS}~\cite{velasco2020knosos} relied on closeness to omnigeneity (i.e., to perfect optimization with respect to neoclassical transport) to model neoclassical transport with a radially local equation. The new version of the code implements and solves the local equations derived in~\cite{dherbemont2021fow}, valid for stellarators of large aspect ratio and arbitrary degree of neoclassical optimization as long as the radial electric field is not too small (see below). The rest of the paper is organized as follows: in section~\ref{SEC_MOT}, the motivation and goals are discussed in detail. Section~\ref{SEC_EQ} presents the equations, whose implementation is described in section~\ref{SEC_IMPL} and employed in the examples of section~\ref{SEC_RES}. In section~\ref{SEC_DKES} we show that, where applicable, the new version of \texttt{KNOSOS} reproduces the results of \texttt{DKES}, being orders of magnitude faster (a comparison of the equations solved by \texttt{KNOSOS} and \texttt{DKES} is made in~\ref{SEC_APP}). In section~\ref{SEC_SBP} we illustrate how, when the radial electric field is small, by retaining the effect of the component of the magnetic drift that is tangent to magnetic surfaces, \texttt{KNOSOS} can describe the superbana-plateau transport regime of stellarators close to omnigeneity. Section~\ref{SEC_SUM} summarizes the results and outlines several planned applications of \texttt{KNOSOS}.

\section{Motivation and goal}\label{SEC_MOT}

\begin{figure}
\center
\includegraphics[angle=0,width=.9\columnwidth]{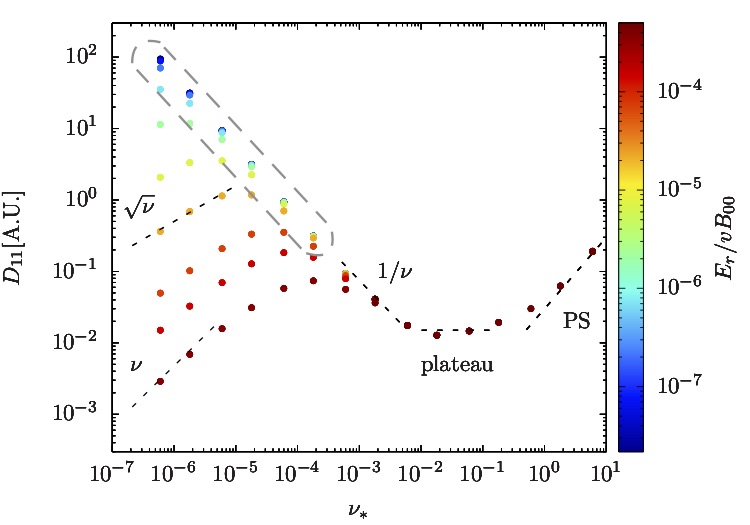}
\caption{Monoenergetic radial transport coefficient $D_{11}$ as a function of the collisionality and normalized radial electric field. The higher collisionality cases have been calculated with \texttt{DKES}, and the lower collisionality ones, with~\texttt{KNOSOS} (in the large aspect ratio limit, as in section~\ref{SEC_DKES}).}
\label{FIG_SKETCH}
\end{figure}

Since the general goal of this work is to provide a fast and accurate calculation of neoclassical transport of stellarators at low collisionality, in this section we provide a brief overview of the relevant transport regimes in the large aspect ratio limit (a more detailed discussion can be found e.g. in \cite{beidler2011ICNTS,calvo2017sqrtnu,calvo2018jpp}). To that end, figure~\ref{FIG_SKETCH} represents the radial transport on a flux-surface of a stellarator configuration as a function of the collisionality $\nu_ *={R\nu}/{(\iota v)}$ and the normalized radial electric field ${E_r}/{(v\fsa{B})}$. Here, $\nu$ is the collision frequency, $R$ the major radius, $\iota$ is the rotational transform, $v$ is the velocity, $E_r$ is the radial electric field and $\fsa{B}$ is the average magnetic field strength on the flux-surface. The level of transport is represented by the so-called monoenergetic transport coefficient $D_{11}$, which is related to the radial energy flux by
\begin{equation}
Q_b \sim \int_0^\infty\mathrm{d} v F_{M,b} \Upsilon_b D_{11}\,.
\end{equation}
A precise definition of $D_{11}$ can be found in~\cite{beidler2011ICNTS,velasco2020knosos}.

At high collisionality, a stellarator configuration displays the same neoclassical transport regimes as an axisymmetric tokamak: Pfirsch-Schl\"uter and plateau. However, due to its intrinsic three-dimensionality, its low-collisionality neoclassical regimes are stellarator specific: for $\epsilon^{3/2}\gg\nu_*\gg \rho_*/\epsilon$, with $\rho_*=\rho/R$ the normalized Larmor radius, $\epsilon=a/R$ the inverse aspect ratio and $a$ the stellarator minor radius, the plasma is in the $1/\nu$ regime; for smaller collisionalities, $\nu_*\ll \rho_*/\epsilon$, the $\sqrt{\nu}$ or $\nu$ regimes dominate transport~\cite{dherbemont2021fow}. All these regimes can be computed with standard neoclassical codes such as \texttt{DKES}~\cite{hirshman1986dkes}. There are, however, two limitations to what codes like \texttt{DKES} can do. 

First, for small $\nu_*$ (and $E_r$ of standard size) the distribution function becomes increasingly localized in phase space. For this reason, while the high collisionality regimes are easy to calculate numerically, the calculation of the $\sqrt{\nu}$ or $\nu$ fluxes comes at a large computational cost. In the case of the $1/\nu$ regime, this limitation has been circumvented by using a code, \texttt{NEO} \cite{nemov1999neo}, that efficiently solves a bounce-averaged drift-kinetic equation specific of the $1/\nu$ regime of stellarators. However, no equivalently fast code exists for the $\sqrt{\nu}$ and $\nu$ regimes for general stellarator geometry (fast codes such as \texttt{GSRAKE}~\cite{beidler1995gsrake,beidler2011ICNTS} and \texttt{NEO-2}~\cite{kernbichler2016neo2} can be applied to particular geometries). The lack of a fast neoclassical code for such regimes has an impact on the availability of stellarator optimization strategies (stellarator optimization is the term employed to describe the tailoring of the magnetic configuration to meet a certain series of criteria). Minimization of neoclassical transport, a standard optimization criterion, usually targets the $1/\nu$ flux~\cite{beidler2021nature}: the transport of electrons (that are in the $1/\nu$ regime) is reduced; this causes a negative $E_r$ that indirectly reduces the transport of ions (that are expected to be in the $\sqrt{\nu}$ or $\nu$ regimes). Although this optimization strategy has proven to be successful (e.g. in the design of the stellarator Wendelstein 7-X), addressing directly the $\sqrt{\nu}$ or $\nu$ flux as well could lead to more efficient approaches, which could be beneficial when additional criteria (such as magnetohydrodinamic stability, reduced turbulence...) exist.

Second, the above-mentioned regimes are obtained by solving a drift kinetic equation that neglects the component of the magnetic drift that is tangent to the flux-surface in the particle orbits~\cite{beidler2011ICNTS}. While this is correct if the aspect ratio $1/\epsilon$ is large and $E_r$ has the standard size ($E_r \sim T/(ae)$, with $T$ the temperature and $e$ the elementary charge; more precise definitions of ``standard size" and ``small" $E_r$ will be provided in sections~\ref{SEC_DKES} and \ref{SEC_SBP}), there exist scenarios in which this drift-kinetic equation is not accurate. Examples of this are compact stellarators (stellarators with values of $1/\epsilon$ down to 2.5 have been designed~\cite{ku2008ariescs}) or plasma scenarios with electron temperature significantly greater than the ion temperature. An extreme example of this is the crossover region in plasmas displaying electron root ($E_r>0$) in the core and ion root ($E_r<0$) closer to the edge, see e.g. \cite{pablant2019ionroot}. In these situations, for $\nu_*\ll \rho_*$, an additional regime may appear~\cite{calvo2017sqrtnu}, the superbanana-plateau regime. However, because of the lack of tangential magnetic drift, codes as such \texttt{DKES} will incorrectly predict a $1/\nu$ regime. The collisionality range where this happens is roughly indicated by a dashed closed line in figure~\ref{FIG_SKETCH}.

The goal of this work is twofold: first, to show that the equations now implemented in \texttt{KNOSOS} produce fast and accurate calculations of neoclassical transport for large aspect ratio stellarators at low collisionality and radial electric field of standard size; second, to provide examples of accurate (provided that the magnetic configuration is close to omnigeneity) computations at low collisionality and small radial electric field.

\section{Equations}\label{SEC_EQ}

We will employ spatial coordinates that are aligned with the magnetic field $\mathbf{B}$. The flux-surfaces are labelled by $\psi=|\Psi_t|$, where $2\pi\Psi_t$ is the toroidal magnetic flux, and $\alpha=\theta-\iota\zeta$ labels magnetic field lines on the surface. Here, $\theta$ and $\zeta$ are Boozer angles and $\iota$ is the rotational transform. Finally, $l$ is the arc length along the magnetic field line. As for velocity phase space, we will employ the magnitude of the velocity $v = |\mathbf{v}|$, the pitch-angle coordinate $\lambda=v_\perp^2/(v^2B)$ and the sign of the parallel velocity $\sigma = v_\parallel/v$. As usually, $v_\parallel=\mathbf{v}\cdot \mathbf{B}/B=\sigma v\sqrt{1-\lambda B}$ and $v_\perp=\sqrt{v^2-v_\parallel^2}$.

In order to compute radial neoclassical transport at low collisionality, we need to calculate, for each bulk species $b$, $g_b(\psi,\alpha,\lambda,v)$: the deviation of the distribution function from a Maxwellian $F_{M,b}$ for trapped particles (i.e. for those whose parallel velocity becomes zero at some point of their orbit). We need to solve the following partial differential equation:
\begin{eqnarray}
\int_{l_{b_1}}^{l_{b_2}} \frac{\mathrm{d}l}{|v_\parallel|} \left(\mathbf{v}_{M,b} + \frac{B}{\fsa{B}}\mathbf{v}_E\right)\cdot\nabla\alpha \left(\partial_\alpha + \partial_\alpha\lambda |_J\partial_\lambda\right) g_b - \int_{l_{b_1}}^{l_{b_2}} \frac{\mathrm{d}l}{|v_\parallel|} C_b^{\mathrm{lin}}[g_b] = \nonumber~~~~~~~~~~~~~~~~~~~~~\\
=-\int_{l_{b_1}}^{l_{b_2}} \frac{\mathrm{d}l}{|v_\parallel|} (\mathbf{v}_{M,b}+\mathbf{v}_E)\cdot\nabla \psi \Upsilon_b F_{M,b}\,,~~~~~~~~~~~~~~~~~~~~~\
\label{EQ_DKE}
\end{eqnarray}
complemented with $g_b=0$ at the boundary between trapped and passing particles. In this equation, the coefficients are integrals over $l$, at constant $\alpha$ and $\lambda$, between bounce points $l_{b_1}$ and $l_{b_2}$, defined as the positions where $v_\parallel=0$:
\begin{equation}
B(\alpha,l_{b_1})=B(\alpha,l_{b_2})=\frac{1}{\lambda}\,.
\end{equation}
On the right-hand side of equation~(\ref{EQ_DKE}), 
\begin{equation}
\Upsilon_b = \frac{\partial_\psi n_b}{n_b} + \frac{\partial_\psi T_b}{T_b}\left(\frac{m_bv^2}{2T_b}-\frac{3}{2}\right)+\frac{Z_be\partial_\psi\varphi_0}{T_b}
\end{equation}
is a combination of thermodynamical forces, 
\begin{equation}
C_b^{\mathrm{lin}}[g_b]=\frac{\nu_{\lambda,b} v_{||}}{v^2 B}\partial_\lambda\left(v_{||}\lambda\partial_\lambda g_b \right)
\end{equation}
 is the linearized pitch-angle collision operator, and
\begin{eqnarray}
\mathbf{v}_{M,b} &=& \frac{m_bv^2}{Z_be}\left(1-\frac{\lambda B}{2}\right)\frac{\mathbf{B}\times\nabla B}{B^3}\nonumber\\
 \mathbf{v}_E &=& -\frac{\nabla\varphi\times\mathbf{B}}{B^2}
\end{eqnarray}
are the magnetic and $E\times B$ drifts. In these equations, as usual, $m_b$ is the mass, $Z_be$ is the charge, $T_b$ is the temperature and $n_b$ is the density. The electrostatic potential $\varphi=\varphi_0(\psi)+\varphi_1(\psi,\alpha,l)$ needs to be estimated by solving the ambipolarity and quasineutrality equations (these equations are written explicitly in \cite{calvo2018jpp,velasco2020knosos} and remain valid for large aspect ratio stellarators) consistently with the drift-kinetic equation. Here, $\varphi_0$ is the piece of $\varphi$ that does not depend on the angular coordinates. Finally, the term with
\begin{equation}
\partial_\alpha\lambda |_J \equiv -\frac{\int_{l_{b_1}}^{l_{b_2}} \frac{\mathrm{d}l}{|v_\parallel|} \lambda\partial_\alpha B}{\int_{l_{b_1}}^{l_{b_2}} \frac{\mathrm{d}l}{|v_\parallel|} B}\label{EQ_CONSJ}
\end{equation}
ensures that the particle orbits conserve the second adiabatic invariant
\begin{eqnarray}
J(\psi,\alpha,v,\lambda)=2v\int_{l_{b_1}}^{l_{b_2}}\mathrm{d}l \sqrt{1-\lambda B}\,.
\end{eqnarray}
This can be checked by employing the identities
 \begin{eqnarray}
\overline{B} &=& - \frac{2}{v^2\tau_b}\partial_\lambda J\nonumber\\
\overline{\partial_\alpha B} &=& - \frac{2}{\lambda v^2\tau_b}\partial_\alpha J\,,
\end{eqnarray}
where the orbit average of a function $f$ is defined as
\begin{eqnarray}
\overline{f}=\frac{1}{v\tau_b}\sum_\sigma \int_{l_{b_1}}^{l_{b_2}}\mathrm{d}l \frac{f(\psi,\alpha,l,v,\lambda,\sigma)}{ \sqrt{1-\lambda B}}
\label{EQ_BAV}
\end{eqnarray}
in the trapped region and
\begin{eqnarray}
\tau_b=\frac{2}{v}\int_{l_{b_1}}^{l_{b_2}}\mathrm{d}l \frac{1}{\sqrt{1-\lambda B}}\,
\end{eqnarray}
is the bounce time.


Let us end this section by laying out the assumptions made for the derivation of equation~(\ref{EQ_DKE}) (details are discussed rigorously in \cite{calvo2017sqrtnu,dherbemont2021fow}). At low collisionality, the motion of trapped particles along $\mathbf{B}$ (i.e. in the $l$ coordinate) is much faster than collisions. In this limit, the distribution function becomes independent of arc-length $l$ and the coefficients of the equation are integrals over $l$ between the bounce points. This is therefore an ansatz that is specific of a bounce-averaged code for radial transport. The collisions in equation~(\ref{EQ_DKE}) are modelled by means of a pitch-angle collision operator, which is common to \texttt{DKES} and all the other \textit{monoenergetic} codes in~\cite{beidler2011ICNTS}, and is valid for trapped particles if the magnetic configuration has large aspect ratio, $\epsilon\ll 1$. 

Finally, there are two limits in which neoclassical transport can be modelled accurately by a drift kinetic equation which only describes trapped particles and that is radially-local (i.e. does not contain terms proportional to $\partial_\psi g_b$, like for example the drift-kinetic equation solved in~\cite{satake2006fortec3d}), such as equation~(\ref{EQ_DKE}).  First, when the configuration is close enough to omnigeneity. The  previous version of~\texttt{KNOSOS}~\cite{velasco2020knosos} made use of this limit. Second,  when, additionally to $\epsilon\ll 1$, $E_r$ is of standard size~\cite{dherbemont2021fow}. In~\ref{SEC_APP} we show that, in this limit, equation~(\ref{EQ_DKE}) is equivalent to the drift-kinetic equation solved by \texttt{DKES}. 

The version of~\texttt{KNOSOS} presented in this work solves numerically equation~(\ref{EQ_DKE}), and is therefore valid in these two limits and in particular can now describe large aspect ratio stellarators irrespectively of its degree of optimization for $E_r$ of standard size. The implementation of equation~(\ref{EQ_DKE}) is discussed in section~\ref{SEC_IMPL} and its application in the two limits is demonstrated in sections~\ref{SEC_DKES} and~\ref{SEC_SBP}, respectively. This discussion and the content of the rest of the paper is outlined in table~\ref{TAB_COMP}.

\begin{table}
\begin{tabular}{c|||c|c}
starting point   & \multicolumn{2}{c}{equation~(\ref{EQ_DKE})} \\
\hline
implemented in & \multicolumn{2}{c}{section~\ref{SEC_IMPL}} \\
\hline
valid for  & \multicolumn{2}{c}{low collisionality, $\epsilon\ll1$ and...} \\
\hline
\hline
\hline
...in two limits & $\mathbf{E_r}$ \textbf{of standard size} & \textbf{closeness to omnigeneity} \\
 & $\mathbf{E_r \sim T_b/(a\,Z_be)}$ & \\
\hline
\hline
\hline
applicability & most stellarator scenarios & applied by all local codes \\
 &  & when $E_r \ll T_b/(a\,Z_be)$ \\
\hline
in this limit &  &  \\
equation~(\ref{EQ_DKE}) leads to & equation~(\ref{EQ_BDKES}) & equation~(\ref{EQ_ODKE}) \\
\hline
derived in & \cite{dherbemont2021fow} & \cite{calvo2017sqrtnu,calvo2018jpp} \\
\hline
employed first in &  \cite{dherbemont2021fow} and this work & \cite{velasco2020knosos}\\
\hline
can describe regimes & $1/\nu$, $\sqrt{\nu}$ and $\nu$ & $1/\nu$, $\sqrt{\nu}$ and superbanana-plateau \\
\hline
main improvement & fast computation  & superbanana-plateau \\
w.r.t. \texttt{DKES} &    &  \\

\hline
relevant for & stellarator optimization & characterization of high $T_i$ \\
& & experiments\\

\hline
exemplified in & section~\ref{SEC_DKES} & section~\ref{SEC_SBP}
\end{tabular}
\caption{Summary of assumptions and applicability of equation~(\ref{EQ_DKE}).}\label{TAB_COMP}
\end{table}

\section{Numerical solution of equation~(\ref{EQ_DKE})}\label{SEC_IMPL}

In this section we provide an overview of how equation~(\ref{EQ_DKE}) is solved. The same implementation will be valid for the two limits in which the equation will be employed. In particular, most of what was discussed in section 3 of~\cite{velasco2020knosos} for stellarators close to omnigeneity remains valid, and in this section we will focus mainly on what has changed. We start by providing a more explicit expression of equation~(\ref{EQ_DKE}):
\begin{eqnarray}
\left(I_{v_{M,\alpha}} (\alpha,\lambda)+\frac{\partial_\psi\varphi_0}{v_{d,b}}I_{v_E,\alpha}(\alpha,\lambda)\right)\left(\partial_\alpha + \partial_\alpha\lambda |_J\partial_\lambda\right) g_b\,\nonumber  \\ -   \frac{\nu_{\lambda,b}}{v_{d,b}} \partial_\lambda\left[ I_\nu(\alpha,\lambda) \partial_\lambda g_b\right]
= - \left( I_{v_{M,\psi}}(\alpha,\lambda)+\frac{1}{v_{d,b}}I_{v_{E,\psi}}(\alpha,\lambda)\right) F_{M,b}\Upsilon_b\,,
\label{EQ_NDKE}
\end{eqnarray}
with
\begin{eqnarray}
I_{v_{E,\alpha}}&=&\Psi_t'\int_{l_{b_1}}^{l_{b_2}} \frac{\mathrm{d}l}{\sqrt{1-\lambda B}}\,,\nonumber\\
I_{v_{M,\alpha}}&=&\int_{l_{b_1}}^{l_{b_2}}\frac{\mathrm{d}l }{\sqrt{1-\lambda B}}\left(1-\frac{\lambda B}{2}\right)\left[\Psi_t'\frac{\partial_\psi B}{B}+
 \frac{B_\zeta\partial_\theta B - B_\theta\partial_\zeta B}{B|B_\zeta+\iota B_\theta|}\zeta\partial_\psi \iota\right]\,,\nonumber\\
I_{v_{E,\psi}}&=&\int_{l_{b_1}}^{l_{b_2}} \frac{\mathrm{d}l}{\sqrt{1-\lambda B}}\frac{B_\theta\partial_\zeta \varphi_1 - B_\zeta\partial_\theta \varphi_1}{|B_\zeta+\iota B_\theta|}\,,\nonumber\\
I_{v_{M,\psi}}&=&\int_{l_{b_1}}^{l_{b_2}} \frac{\mathrm{d}l }{\sqrt{1-\lambda B}}\left(1-\frac{\lambda B}{2}\right)\frac{B_\theta\partial_\zeta B - B_\zeta\partial_\theta B}{B|B_\zeta+\iota B_\theta|}\,,\nonumber\\
I_\nu&=&\int_{l_{b_1}}^{l_{b_2}} \mathrm{d}l\frac{\lambda\sqrt{1-\lambda B}}{B}\,,\nonumber\\
\partial_\alpha\lambda |_J  &=& -\left(\int_{l_{b_1}}^{l_{b_2}}\mathrm{d}l \frac{B}{\sqrt{1-\lambda B}}\right)\Big/ \left(\int_{l_{b_1}}^{l_{b_2}}\mathrm{d}l \frac{\lambda\partial_\alpha B}{\sqrt{1-\lambda B}}\right)\,.
\label{EQ_BINT}
\end{eqnarray}
Here, $B_\psi=0$, $B_\theta$ and $B_\zeta$ are the covariant components of $\mathbf{B}$ in Boozer coordinates
\begin{eqnarray}
\mathbf{B}=B_\psi\nabla\psi+B_\theta\nabla\theta+B_\zeta\nabla\zeta\,,
\end{eqnarray}
with $B_\psi=0$ in the low-$\beta$ approximation. 

\begin{figure}
\centerline{\includegraphics[angle=0,width=0.8\columnwidth]{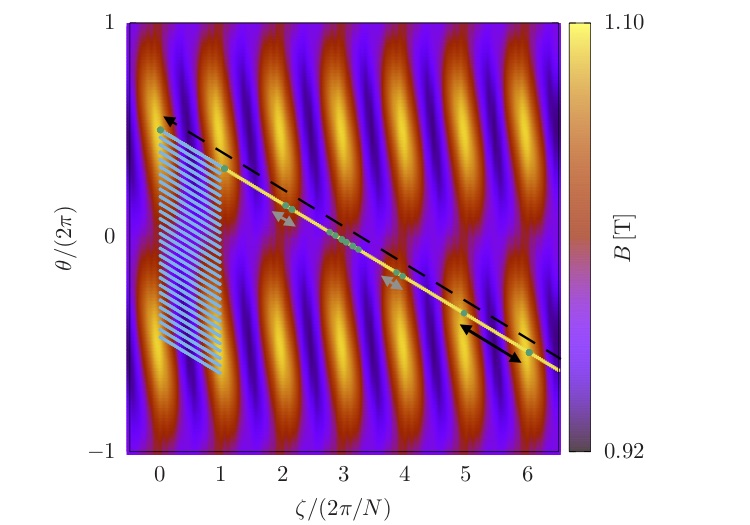}}
\centerline{\includegraphics[angle=0,width=0.8\columnwidth]{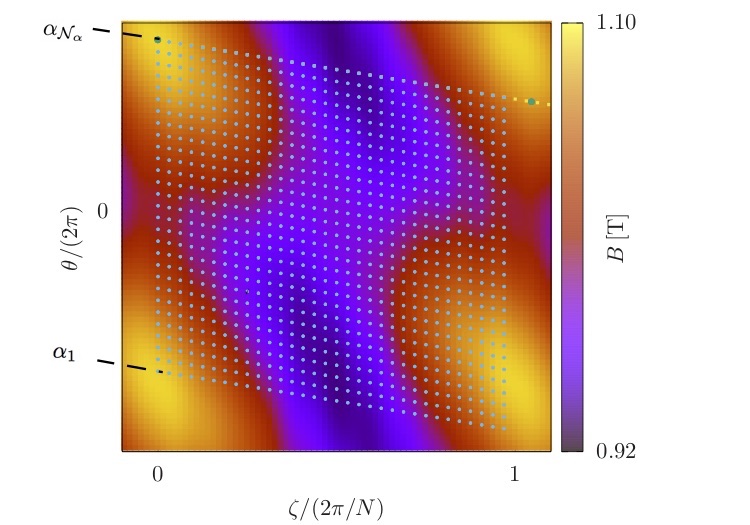}}
\caption{Construction of the angular grid (see text) for a flux surface of Wendelstein 7-X (top); zoom (bottom).}
\label{FIG_GRID}
\end{figure}

The radial coordinate $\psi$ is a parameter in equation~(\ref{EQ_BINT}), and radially-local independent calculations can thus be performed at different flux-surfaces of a magnetic configuration. Similarly, the magnitude of the velocity $v$ is a parameter, and so-called \textit{monoenergetic}~calculations can be performed separately for a range of values of $v$; then, quantities such as the radial energy flux can be computed efficiently by Gauss-Laguerre integration on $v$ of the corresponding solutions, as discussed at the end of section 3.3 of~\cite{velasco2020knosos}.  

The main difference between \texttt{KNOSOS} and other neoclassical codes is that $l$ is not a variable: $g_b$ is independent of the arc--length and the coefficients of equation~(\ref{EQ_BINT}) are bounce-integrals, whose fast and accurate integration by means of an extended midpoint rule is discussed in detail in section 3.2 and the appendices of~\cite{velasco2020knosos}. Furthermore, the dependence of transport on the magnetic configuration and the species has been separated in equation~(\ref{EQ_BINT}). The integrals of equation (\ref{EQ_BINT}) depend on the former, while the latter is encapsulated in 
\begin{eqnarray}
v_{d,b}&\equiv&\frac{m_bv^2}{Z_be}\,,
\end{eqnarray}
together with  $\nu_{\lambda,b}$, $F_{M,b}$ and $\Upsilon_b$ (these three quantities contain as well information of the kinetic profiles). For this reason, the integrals of equation~(\ref{EQ_BINT}) need to be calculated once and can then be employed in subsequent calculations for other species or plasma parameters.

The resulting drift-kinetic equation is thus a differential equation in two variables, $\alpha$ and $\lambda$. In the next two subsections, we describe how the bidimensional grid is constructed, and how the drift-kinetic equation is discretized.

\subsection{New spatial grid}\label{SEC_GRID}

The distribution function $g_b$, as well as the coefficients that need to be computed to solve the drift-kinetic equation, are evaluated in a discrete set of points ($\alpha,\lambda$). In this subsection we summarize how these points are selected. The pitch angle variable is discretized exactly as in~\cite{velasco2020knosos}: we employ a uniform grid with ${\cal{N}}_\lambda+1$ values between $\lambda_1\equiv 1/B_{max}$ and $\lambda_{{\cal{N}}_\lambda+1}\equiv 1/B_{min}$, being $B_{max}$ and $B_{min}$ the maximum and minimum of $B$ on the flux-surface respectively. $\lambda_{{\cal{N}}_\lambda+1}$ will be a \textit{ghost} point where the boundary conditions at the bottom of the well are imposed. It must be noted that, in a non-omnigenous stellarator, the magnetic field well depth depends on $\alpha$, an so does the maximum reachable value of $\lambda$. For an arbitrary field line there may exist less points than $\lambda_{{\cal{N}}_\lambda+1}$ in the $\lambda$ grid (when this is the case, the role of ghost point will be played by a grid point with smaller index). As in~\cite{velasco2020knosos}, when integrating in $\lambda$, we will use the extended trapezoidal rule.

We use figure~\ref{FIG_GRID} (top), which shows one example of stellarator flux surface (of the standard Wendelstein 7-X configuration, discussed in section~\ref{SEC_DKES}) to describe how the angular grid is now built. First, a field line is followed until it completes a large number (of the order of 100) of toroidal periods, ${\cal{N}}_\alpha$. Along this line, the yellow points are selected with uniform spacing in the toroidal angle, starting at $\zeta=0$. This spacing is $(2\pi/N)/{\cal{N}}_l$, being ${\cal{N}}_l$ a power of 2 comparable to ${\cal{N}}_\alpha$ (this will be useful for a fast computation of the Fourier transform, needed when solving quasineutrality). Next, periodicity is employed to project all these grid points onto the first toroidal period. The result is a bidimensional grid in $\alpha$ and $l$, of size ${\cal{N}}_\alpha\times{\cal{N}}_l$, as shown in figure~\ref{FIG_GRID} (bottom) in light blue. It is aligned with the field lines and (except for values of $\iota$ close to a rational surface) samples correctly a toroidal period (in general, non uniformly in $\alpha$). This is a procedure similar to the one employed e.g. by~\texttt{NEO}~\cite{nemov1999neo}. In the previous version of~\texttt{KNOSOS}, several field lines, uniformely distributed in $\alpha$, were followed, instead of one. 

This new discretization facilitates an improved sampling of the region of phase space where barely trapped particles live, which can be necessary for an accurate calculation of the $\sqrt{\nu}$ and $\nu$ regimes at low collisionality. The green circles of figure~\ref{FIG_GRID} correspond to maxima of the magnetic field strength along the magnetic field line. Most trapped particles are so in a major well that coincides with one field period (black continuous arrow). In other words, their bounce points $l_{b_1}$ and $l_{b_2}$ are two consecutive green points, separated toroidally by a characteristic angular distance $\sim 2\pi/N$. This distance will be smaller for large values of $\lambda$, close to the bottom of the magnetic well. Conversely, particles with relatively small $\lambda$ will have their bounce points separated by several toroidal periods. The field-line segment that constitutes the spatial grid has been chosen so that the value of $B$ at the two extremes is as close as possible to $B_{max}$ (in a non-rational flux-surface, $B$ at the two extremes could be exactly equal to $B_{max}$ only after an infinite number of toroidal turns). As a consequence of this, those particles with $\lambda$ close enough to $1/B_{max}$ will bounce back and forth between the two extremes of the segment (part of the trajectory is marked by the black dashed arrow). These particles are trapped with their bounce points separated ${\cal{N}}_\alpha$ toroidal field periods. If we have chosen a long enough segment, the trajectory practically samples the whole flux-surface, and we can set this $\lambda$ to be the boundary between trapped and passing, i.e., $g_b=0$ for these particles. Particles with smaller $\lambda$ will be considered to be passing, and will not contribute to radial transport.

To conclude this subsection, we note that several ripple wells, indicated by grey arrows, have been found in the example of figure~\ref{FIG_GRID} (grey arrows). As a consequence of this, at a given $\alpha$ and $\lambda$, several wells may exist (i.e., several pairs of $l_{b_1}$ and $l_{b_2}$), and an additional integer label $w$ needs to be employed for them. Altogether, three integers can be used to label any point $(\alpha_i,\lambda_j,w)$ of this grid: $i$ runs from 1 to  ${\cal{N}}_\alpha$,  $j$ from 1 to ${\cal{N}}_\lambda$ and $w=I,II...$ is an integer that labels wells for a given $\alpha$ and $\lambda$. At a given point, we define $g_{i,j,w}\equiv g_b(\alpha_i,\lambda_j,w)$,  $I_{\nu,i,j,w}\equiv I_\nu(\alpha_i,\lambda_j,w)$, and the same applies to any function of the phase space (in order to ease the notation, $g_{i,j,w}$ does not contain a species index).

\subsection{New discretization of the orbits}\label{SEC_SOLDKE}

In section~\ref{SEC_GRID} we have built the grid in variables $\alpha$ and $\lambda$.  The final step in the discretization of the drift-kinetic equation is how we approximate the derivatives of $g_b$ of equation~(\ref{EQ_NDKE}) at each point of this grid.

Three different kinds of terms need to be discretized: the boundary conditions, the collision operator and the orbits. The two former contain only derivatives in $\lambda$ at fixed $\alpha$, and have not changed with respect to~\cite{velasco2020knosos}, where they were explained in detail. Things are different with the orbits, since both the drift-kinetic equation and the grid are now slightly different. This term reads
\begin{eqnarray}
\left(I_{v_{M,\alpha}} (\alpha,\lambda)+\frac{\partial_\psi\varphi_0}{v_{d,b}}I_{v_E,\alpha}(\alpha,\lambda)\right)\left(\partial_\alpha + \partial_\alpha\lambda |_J\partial_\lambda\right) g_b\,.
\label{EQ_DALPHA}
\end{eqnarray}
As written in equation (\ref{EQ_DALPHA}), this term contains derivatives in $\alpha$ (at constant $\lambda$) and $\lambda$ (at constant $\alpha$) and, at an arbitrary point, non-centered finite differences with first-order accuracy are used. For a given flux surface, for each solution of the drift-kinetic equation, the sign of the coefficient in front of $\partial_\alpha g_b$ (i.e. the direction of the flow in the $\alpha$ direction) indicates whether forward or backward differences are employed, i.e.
\begin{equation}
\partial_\alpha g_b|_{i,j,w}=\frac{g_{i+1,j,w}-g_{i,j,w}}{\alpha_{i+1}-\alpha_i}
\label{EQ_FDALPHA}
\end{equation}
 or
 \begin{equation}
\partial_\alpha g_b|_{i,j,w}=\frac{g_{i,j,w}-g_{i-1,j,w}}{\alpha_{i}-\alpha_{i-1}}\,.
\label{EQ_BDALPHA}
\end{equation}
The same applies to $\partial_\lambda g_b$, where forward differences
\begin{equation}
\partial_\lambda g_b|_{i,j,w}=\frac{g_{i,j+1,w}-g_{i,j,w}}{\lambda_{j+1}-\lambda_{j}}
\label{EQ_FDLAMBDA}
\end{equation}
 or backward differences 
 \begin{equation}
\partial_\lambda g_b|_{i,j,w}=\frac{g_{i,j,w}-g_{i,j-1,w}}{\lambda_{j}-\lambda_{j-1}}
\label{EQ_BDLAMBDA}
\end{equation}
are used depending on the sign of the flow and of $\partial_\alpha\lambda |_J$.  The direction of the flow depends on the plasma parameters, species, and $\lambda$. For this reason, during the first stages of the neoclassical calculation, all the above discretizations are calculated. Then, for any particular subsequent calculation (for instance, ions in the presence of a given value of $E_r$) and for each $\lambda$, the appropriate combination of the pre-calculated discretizations is employed at different points of phase-space.

\begin{figure}
\centering
\includegraphics[angle=0,width=0.8\columnwidth]{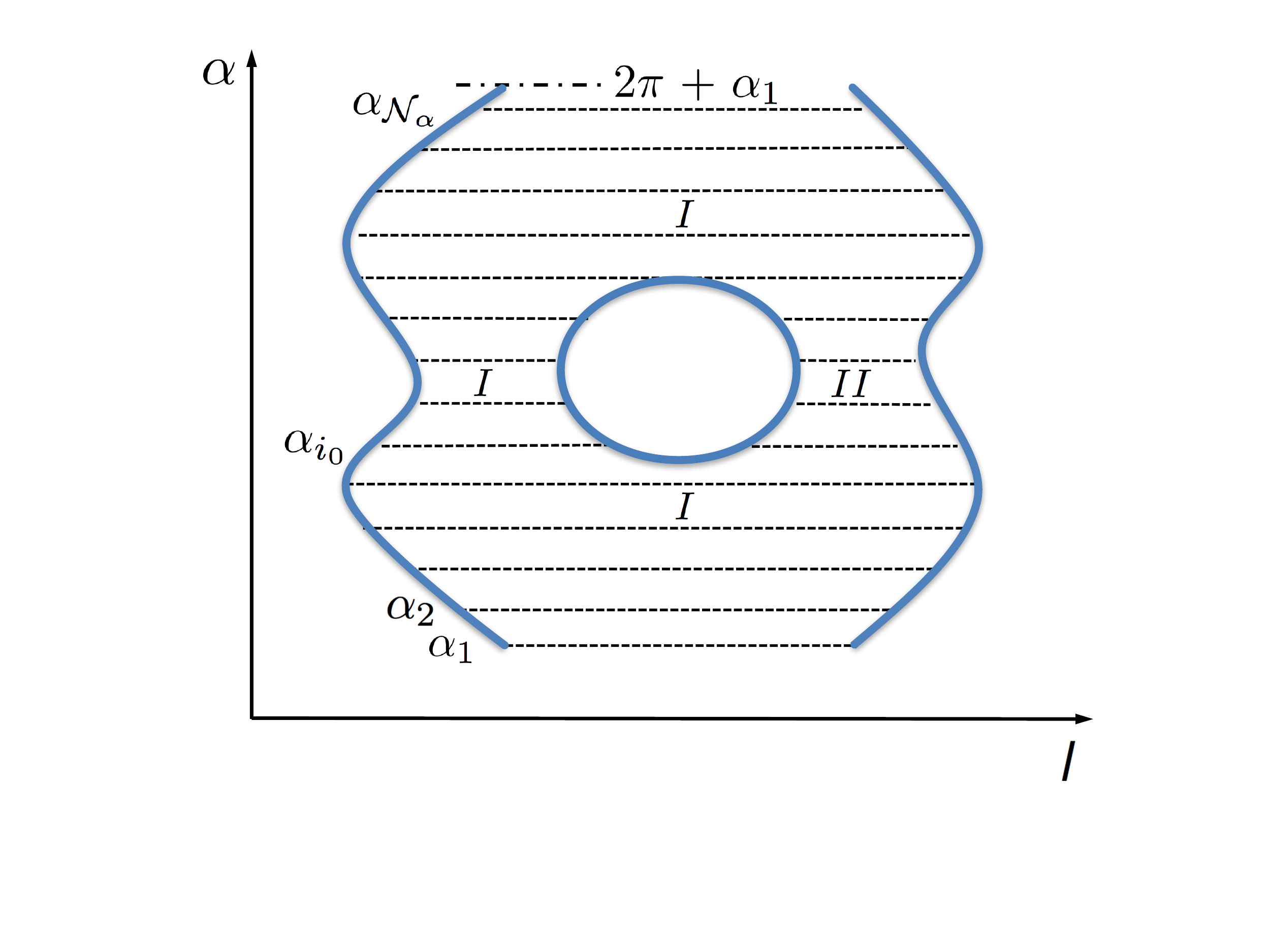}\vskip-1.5cm
\includegraphics[angle=0,width=0.8\columnwidth]{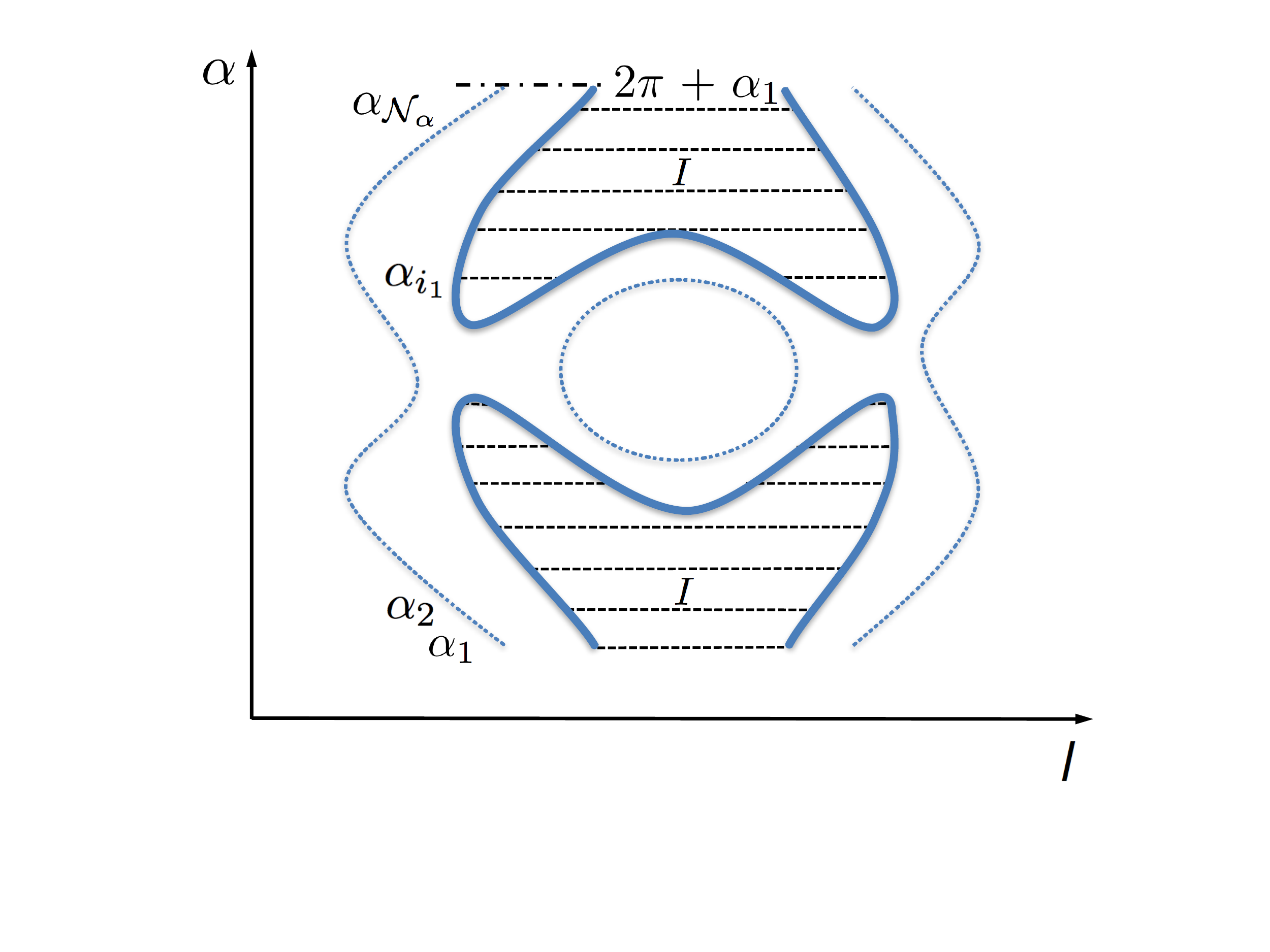}\vskip-0.5cm
\caption{Top: sketch of grid in $\alpha$ space at fixed $\lambda$. The derivatives with respect to $\alpha$ at constant $\lambda$ are discretized as in equations~(\ref{EQ_FDALPHA}) and (\ref{EQ_BDALPHA}) except close to the limits of the grid ($\alpha_1$ and $\alpha_{{\cal{N}}_\alpha}$) and to bifurcations (e.g. $\alpha_{i_0}$). Bottom:  sketch of grid in $\alpha$ space at larger $\lambda$ (the grid at smaller $\lambda$ is plotted for reference in dashed thin blue line). {$\alpha_{i_1}$} is a point where the backward derivative is discretized as discussed in equation~(\ref{EQ_BDALPHAJ}).}
\label{FIG_ALPHA}
\end{figure}

An sketch of the $\alpha$ grid is depicted in figure~\ref{FIG_ALPHA} (top). In this example, for small and large values of $\alpha$, equations (\ref{EQ_FDALPHA}) and (\ref{EQ_BDALPHA}) can be employed (except at $\alpha_1$ and $\alpha_{{\cal{N}}_\alpha}$, where periodicity needs to be imposed, see below). However, things are different in the vicinity of $\alpha_{i_0}$. When moving from small values of $\alpha$ up to $\alpha_{i_0}$, $B$ becomes larger than $1/\lambda_j$ for some values of $l$. As a consequence of this, the orbit \textit{breaks} into two orbits, one in region I and the other one in region II. Continuity of $g_b$ is imposed there to discretize the forward derivative at $i_0-1$,
\begin{eqnarray}
\partial_\alpha g_b|_{i_0-1,j,I}=\frac{g_{i_0,j,I}-g_{i_0-1,j,I}}{\alpha_{i_0}-\alpha_{i_0-1}}=\frac{g_{i_0,j,II}-g_{i_0-1,j,I}}{\alpha_{i_0}-\alpha_{i_0-1}}\,,
\label{EQ_DALPHAB1}
 \end{eqnarray}
 and the backward derivative at $i_0$,
 \begin{eqnarray}
\partial_\alpha g_b|_{i_0,j,I}=\frac{g_{i_0,j,I}-g_{i_0-1,j,I}}{\alpha_{i_0}-\alpha_{i_0-1}}\,,\nonumber\\
\partial_\alpha g_b|_{i_0,j,II}=\frac{g_{i_0,j,II}-g_{i_0-1,j,I}}{\alpha_{i_0}-\alpha_{i_0-1}}\,.
\label{EQ_DALPHAB2}
 \end{eqnarray}
At values of  $\alpha$ immediately above  $\alpha_{i_0}$, equations (\ref{EQ_FDALPHA}) and (\ref{EQ_BDALPHA})  are again valid (and applied separately at regions I and II). However, there is a value of $\alpha$, at which regions I and II merge again into a single well labelled again $I$. There, expressions equivalent to equations (\ref{EQ_DALPHAB1}) and (\ref{EQ_DALPHAB2}) need to be employed.

Periodicity in $\alpha$ is easily imposed by replacing equation (\ref{EQ_FDALPHA}) at $i={\cal{N}}_\alpha$ with
\begin{equation}
\partial_\alpha g_b|_{{\cal{N}}_\alpha,j,w}=\frac{g_{1,j,w}-g_{{\cal{N}}_\alpha,j,w}}{2\pi+\alpha_{1}-\alpha_{{\cal{N}}_\alpha}}\,,
\label{EQ_FDALPHAP}
\end{equation}
and equation (\ref{EQ_BDALPHA}) at $i=1$ with
\begin{equation}
\partial_\alpha g_b|_{1,j,w}=\frac{g_{1,j,w}-g_{{\cal{N}}_\alpha,j,w}}{2\pi+\alpha_{1}-\alpha_{{\cal{N}}_\alpha}}\,.
\end{equation}
Close to the boundary between passing and trapped, at $j=2$, equation (\ref{EQ_BDLAMBDA}) needs to be complemented with the boundary condition 
\begin{equation}
g_b|_{i,1,w}=0\,,
\end{equation}
which yields
\begin{equation}
\partial_\lambda g_b|_{i,2,w}=\frac{g_{i,2,w}}{\lambda_{2}-\lambda_{1}}\,.
\end{equation}
At the bottom, for the field line where $B$ reaches the minimum value on the flux-surface, the regularity condition reads
\begin{equation}
\partial_\lambda\left[ I_\nu \partial_\lambda g_b\right]|_{i,{\cal{N}}_\lambda,w} = -I_{\nu,i,{\cal{N}_\lambda}-1,w}\frac{g_{i,{\cal{N}}_\lambda,w}-g_{i,{\cal{N}}_\lambda-2,w}}{(\lambda_{{\cal{N}}_\lambda+1}-\lambda_{{\cal{N}}_\lambda-1})(\lambda_{{\cal{N}}_\lambda}-\lambda_{{\cal{N}}_\lambda-2})}\,,
\label{EQ_BOTTOM}
\end{equation}
where we have employed that $\partial_\lambda g_b$ is zero at exactly the bottom of the well, $j=\cal{N}_\lambda+1$, and $g_{i,{\cal{N}}_\lambda+1,w}$ (which is defined on a region of phase-space of vanishing volume) does not need to be computed. Additionally, because $\partial_\lambda\left[ I_\nu \partial_\lambda g_b\right]|_{i,{\cal{N}}_\lambda,w}$ is not written explicitly in terms of a second order derivative, its discretization extends beyond nearest neighbours. For field lines where the minimum of $B$ is larger, similar contour conditions can be written, that avoid having to evaluate equation (\ref{EQ_FDLAMBDA}) at points $j$ such that $g_{i,j+1,w}$ is not defined.

This problem is not restricted to the bottom of the well or to the derivative with respect to $\lambda$. While in an omnigenous magnetic field, the contours of minimum $B$ on a flux surface encircle (toroidally, poloidally, or helically) the plasma, this is not  the case for a generic stellarator, where local minima of $B$ exist on the flux surface. Close to these minima, moving in $\alpha$ at constant large $\lambda$ is not always possible, as these trajectories may not exist, and the same happens with moving in $\lambda$ at constant $\alpha$ even if  $j>{\cal{N}}_\lambda-1$. This situation is illustrated in figure~\ref{FIG_ALPHA} (bottom), at $\alpha_{i_1}$, where equation (\ref{EQ_BDALPHA}) cannot be employed. A model was used in~\cite{velasco2020knosos} for dealing with these points, but here we can do that rigorously by recalling that equation (\ref{EQ_DALPHA}) can be interpreted as a derivative in $\alpha$ at constant $J$. Then, one can use equation (\ref{EQ_BDALPHA}) but replacing $g_{i_1-1,j,w}$ with a linear interpolation 
 \begin{equation}
\partial_\alpha g_b|_{i_1,j,w}=\frac{g_{i_1,j,w}-W g_{i_1-1,j_0+1,w}-(1-W)g_{i_1-1,j_0,w}}{\alpha_{i_1}-\alpha_{i_1-1}}\,,
\end{equation}
where $j_0$ and $W$ in the linear interpolation have been chosen so that
 \begin{equation}
J(\alpha_{i_1-1},\lambda_j,w)=WJ(\alpha_{i_1-1},\lambda_{j_0+1},w)+(1-W)J(\alpha_{i_1-1},\lambda_{j_0},w)\,
\label{EQ_BDALPHAJ}
\end{equation}
and, if possible,
 \begin{equation}
J(\alpha_{i_1-1},\lambda_{j_0+1},w)<J(\alpha_{i_1-1},\lambda_j,w)<J(\alpha_{i_1-1},\lambda_{j_0},w)\,
\end{equation}
in order to favour interpolation over extrapolation (this means that, although it will be frequent that $j_0=j$ or $j_0+1=j$, it will not always be the case). This procedure is followed in all the situations in which the nearest neighbour (in the sense of equations (\ref{EQ_FDALPHA}), (\ref{EQ_BDALPHA}),  (\ref{EQ_FDLAMBDA}),  (\ref{EQ_BDLAMBDA})) does not exist.

We end this section by reminding the reader that, for each of the species $b$, we end up with an equation that is linear in $g_b$ and can be written as a linear problem in matrix form. The matrix that represents the orbits and the collision operator is square with approximately ${\cal{N}}_\lambda\times {\cal{N}}_\alpha$ elements per row, and sparse. Although their relative weight varies with $\nu_{\lambda,b}$, $v_{d,b}$ and $\partial_\psi\varphi_0$, the non-zero elements are always at the same position for a given flux surface. As~\cite{velasco2020knosos}, we solve the linear problem with a direct solver from the {\ttfamily PETSc} library based on LU factorization.

\begin{figure}[!t]
\includegraphics[angle=0,width=0.5\columnwidth]{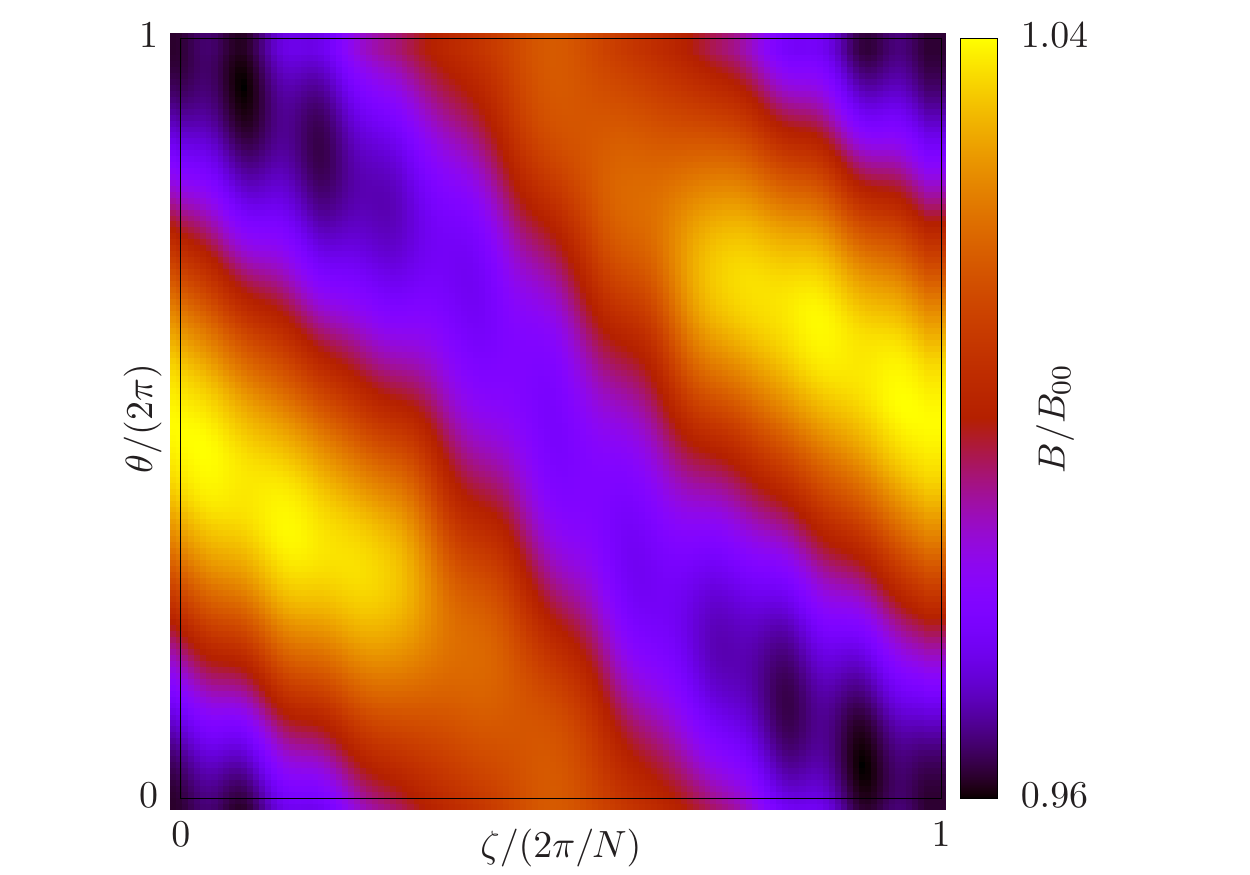}
\includegraphics[angle=0,width=0.5\columnwidth]{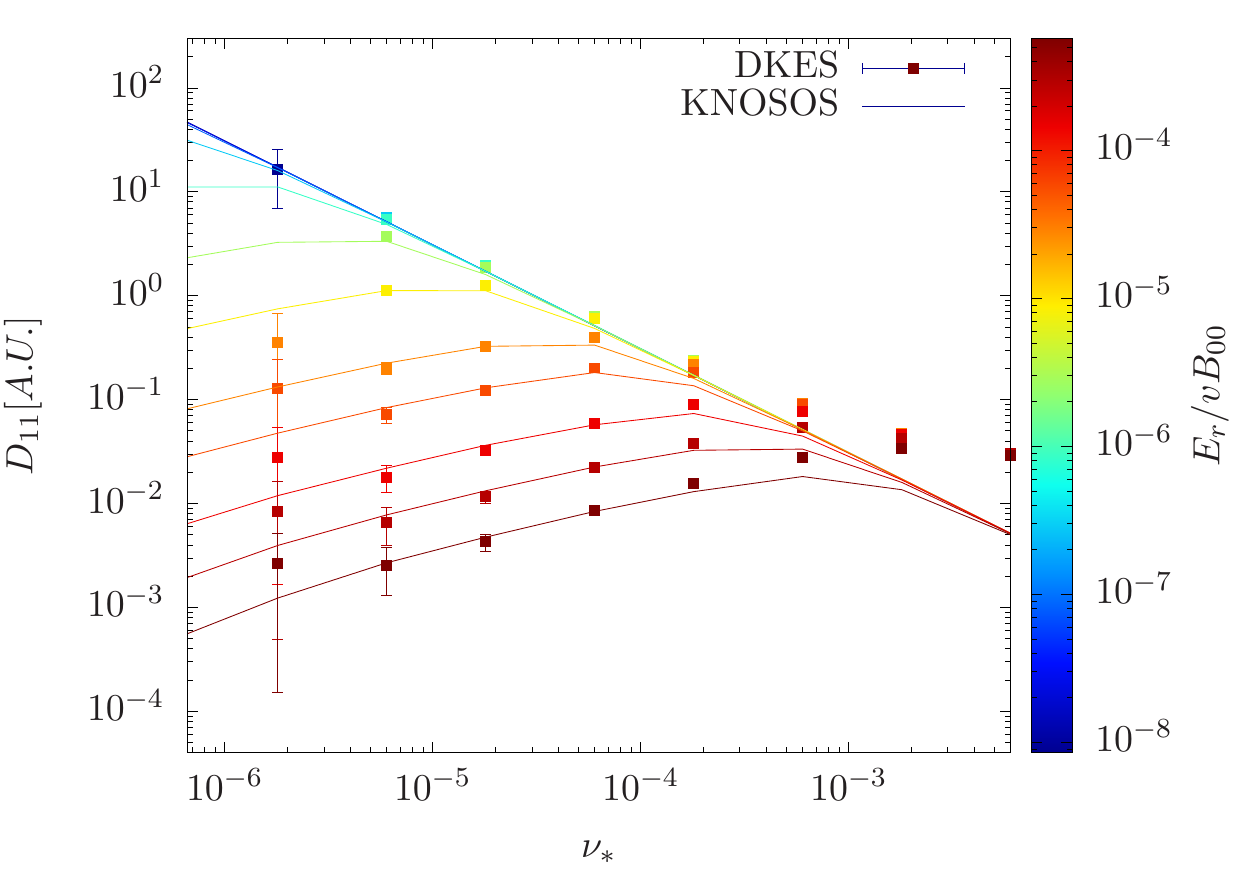}\\
\includegraphics[angle=0,width=0.5\columnwidth]{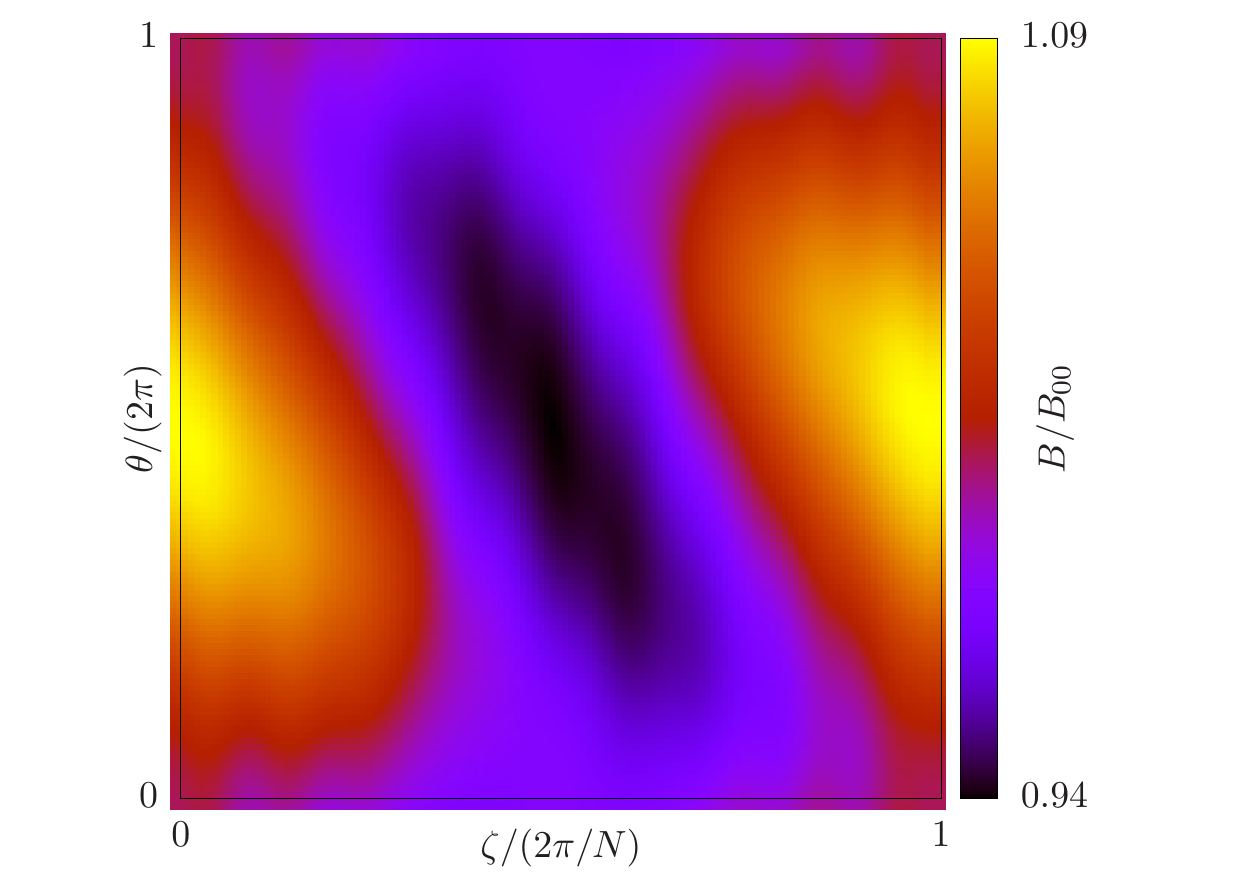}
\includegraphics[angle=0,width=0.5\columnwidth]{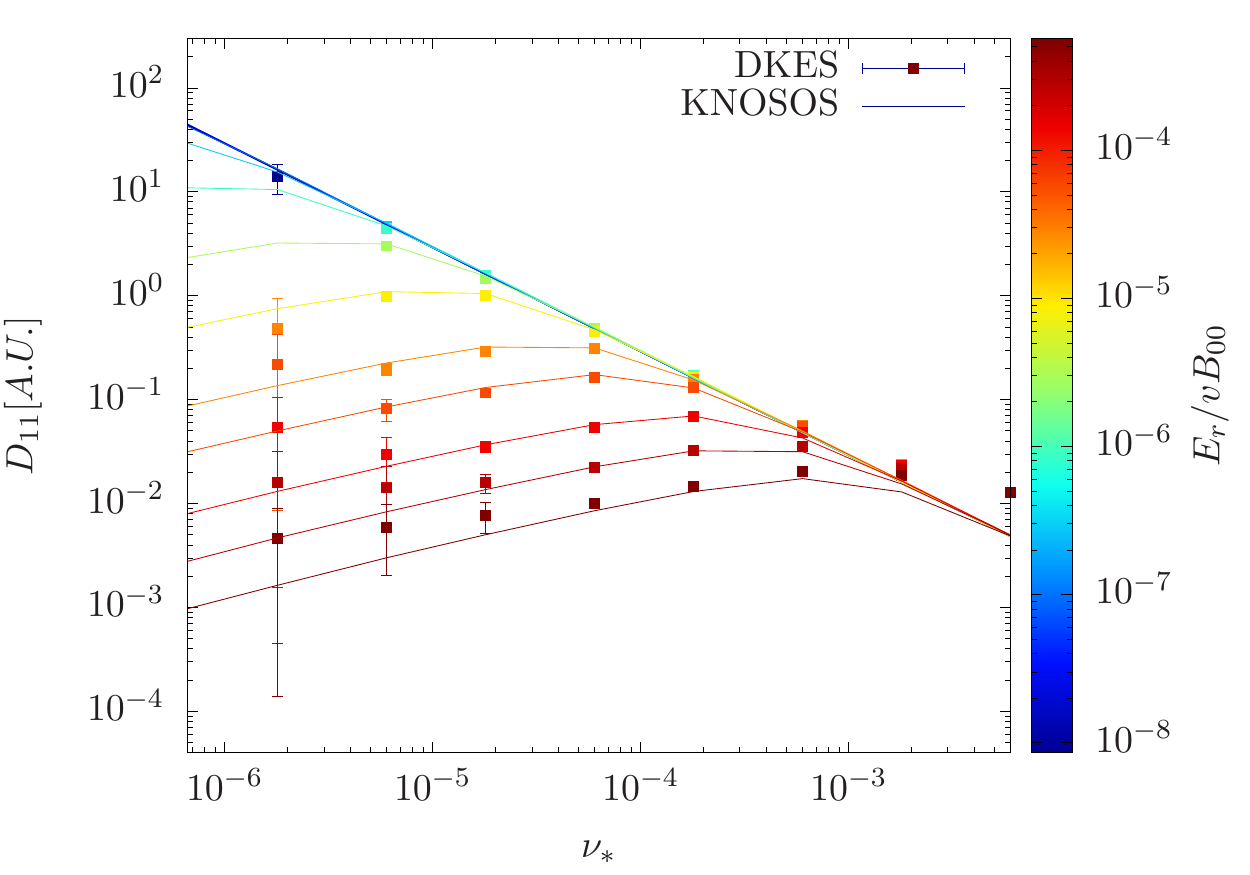}\\
\includegraphics[angle=0,width=0.5\columnwidth]{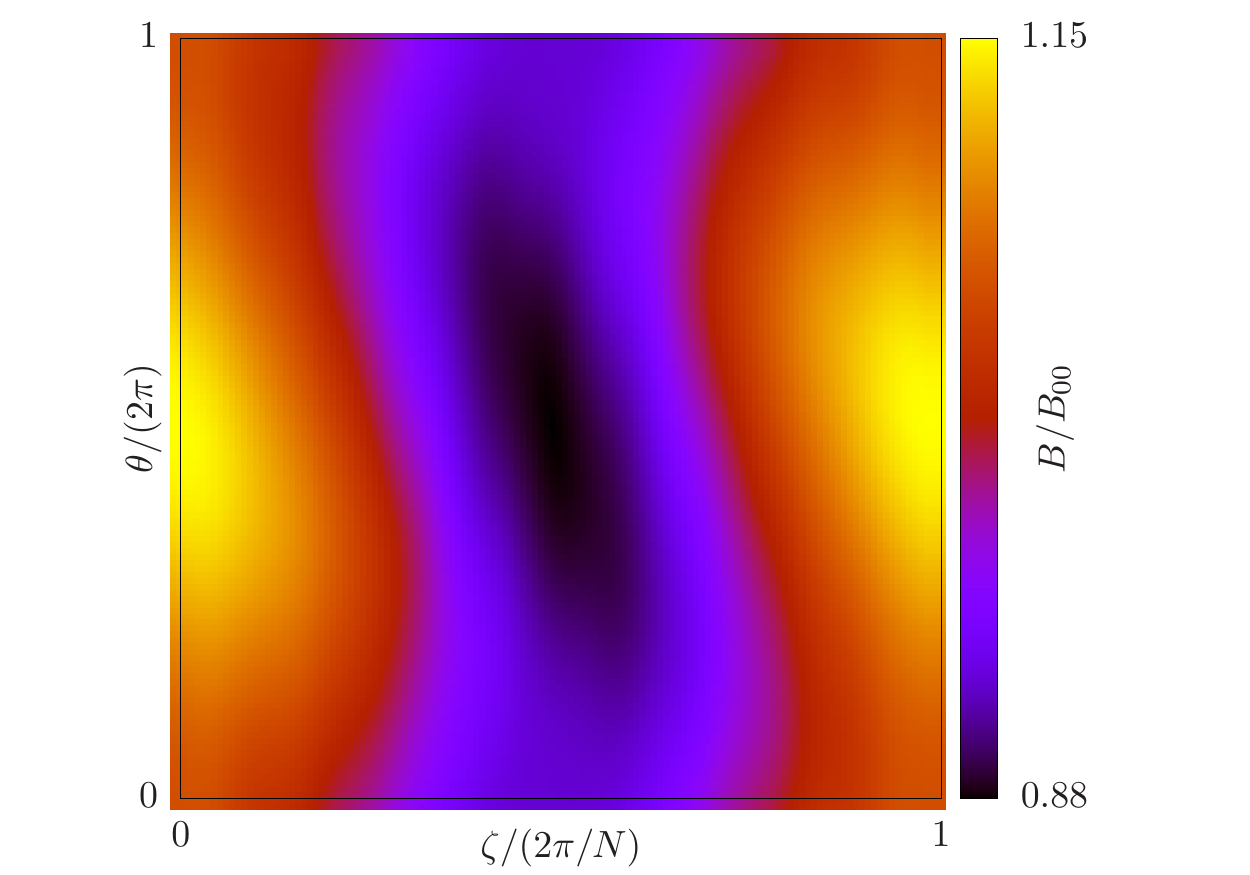}
\includegraphics[angle=0,width=0.5\columnwidth]{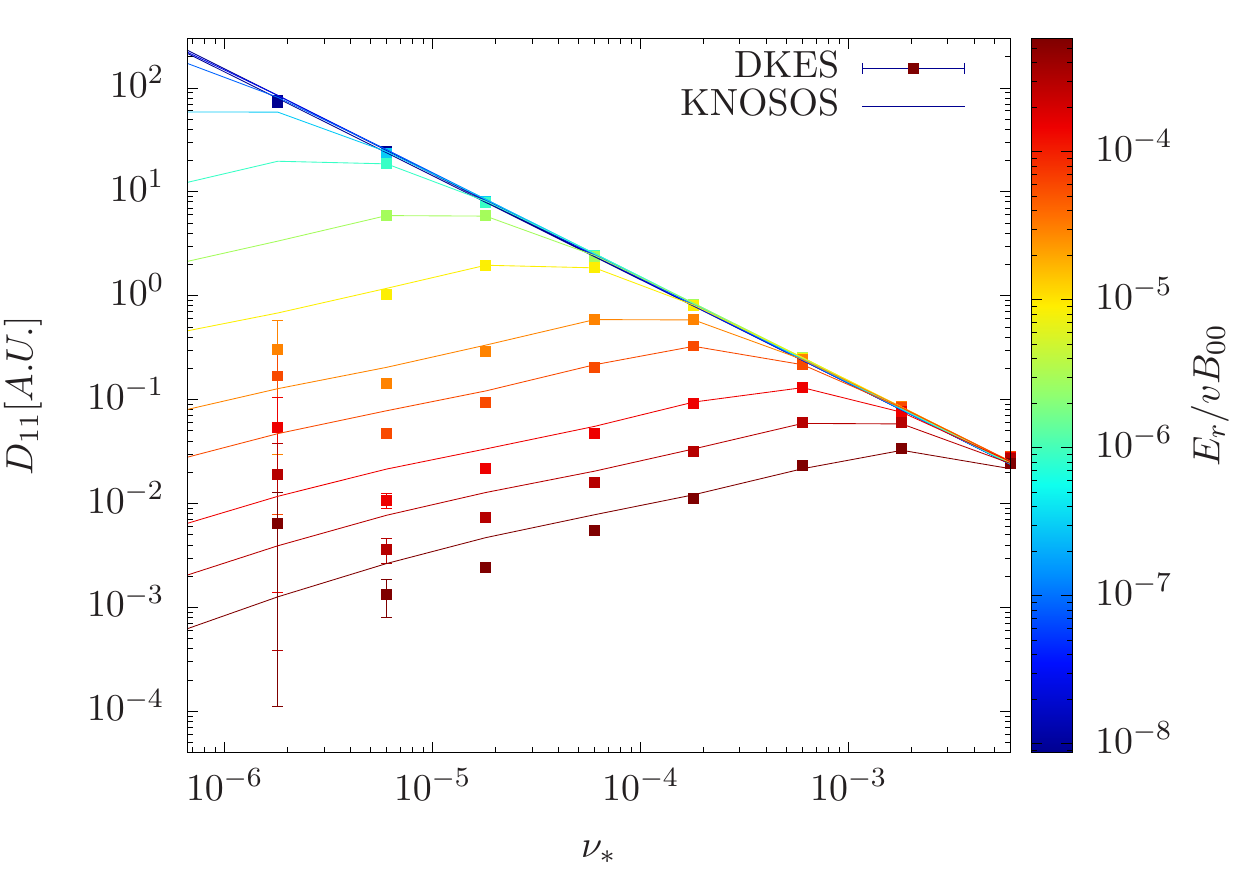}
\caption{Magnetic field strength on the flux-surface (left) and monoenergetic radial transport coefficient $D_{11}$ as a function of the collisionality and normalized radial electric field (right). The calculations correspond to configurations AIM (top), EIM (center) and KJM (bottom) of Wendelstein 7-X.}
\label{FIG_D11b}
\end{figure}

\begin{figure}[!ht]
\includegraphics[angle=0,width=0.5\columnwidth]{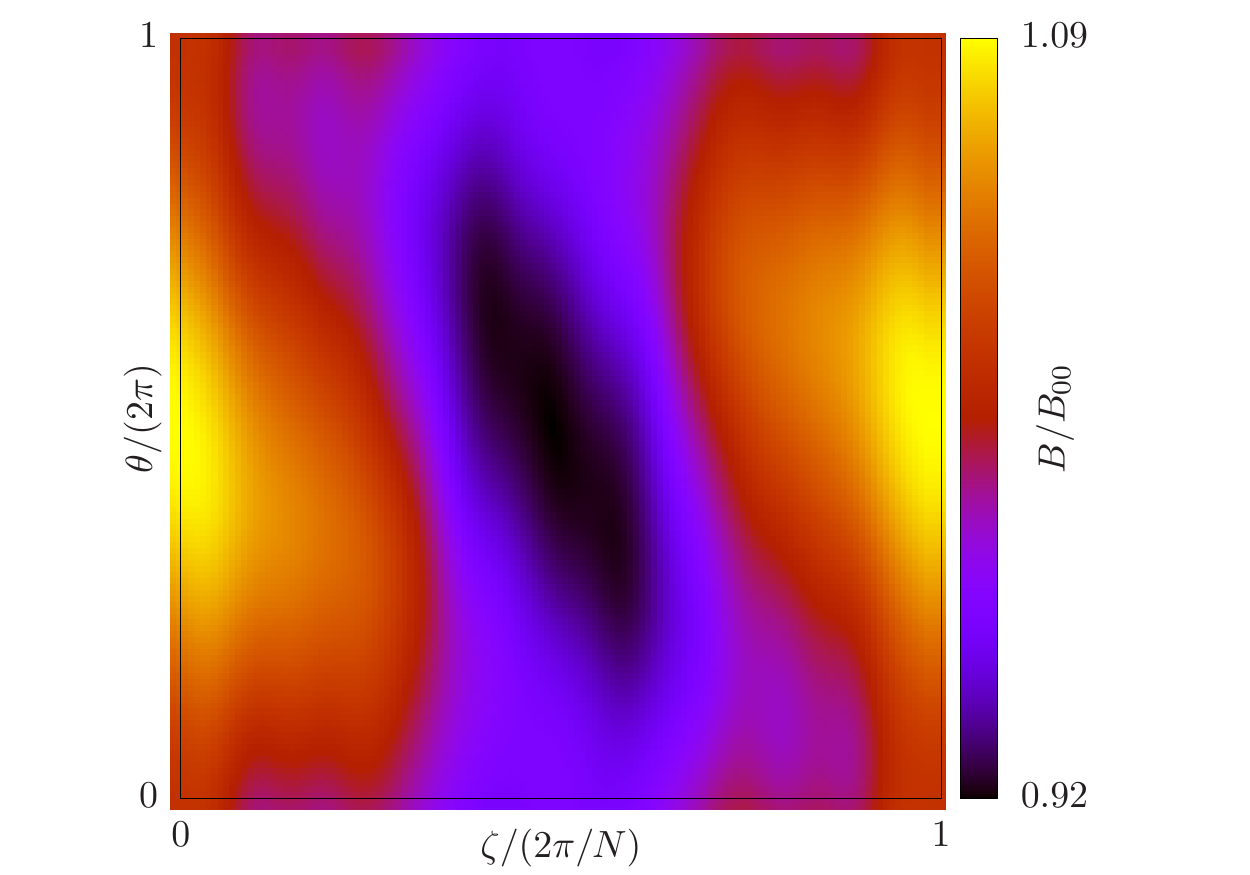}
\includegraphics[angle=0,width=0.5\columnwidth]{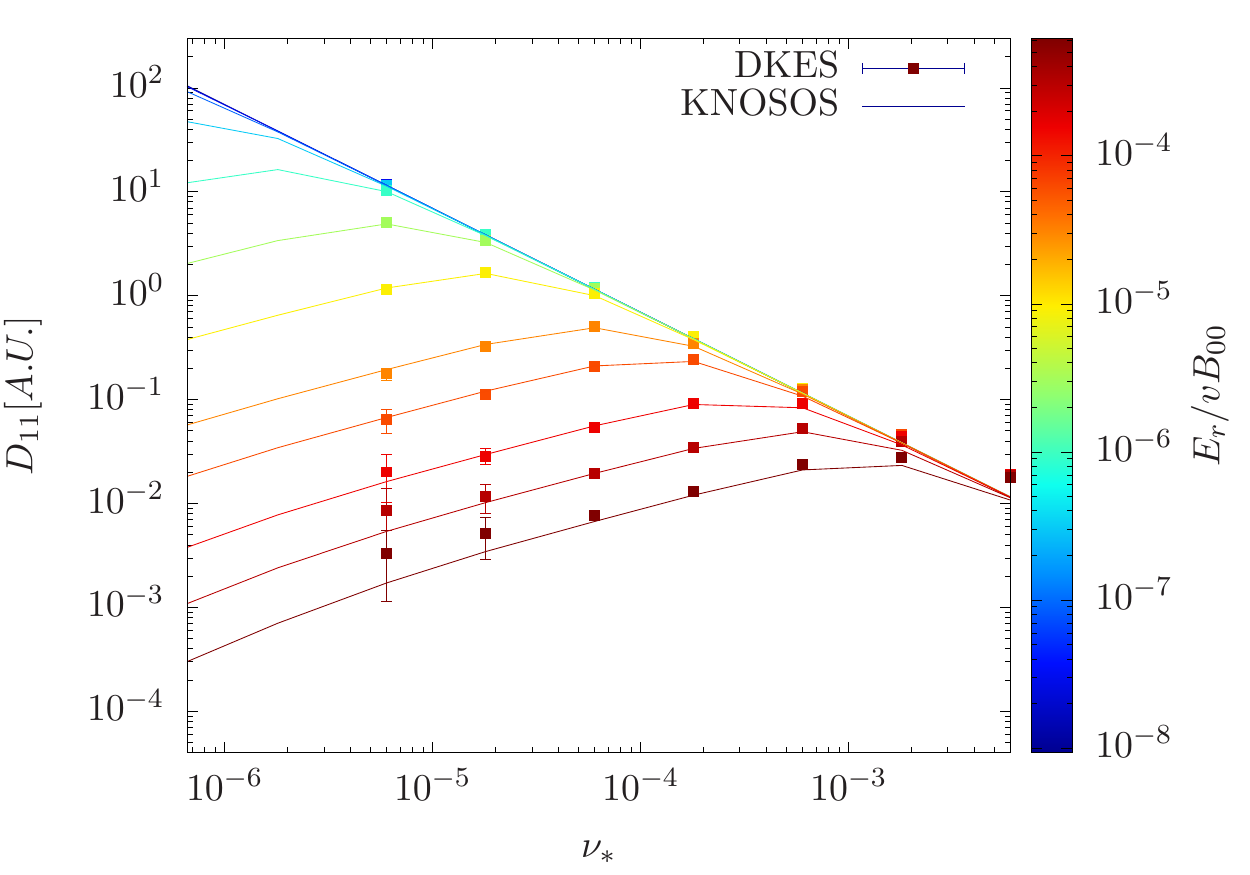}\\
\includegraphics[angle=0,width=0.5\columnwidth]{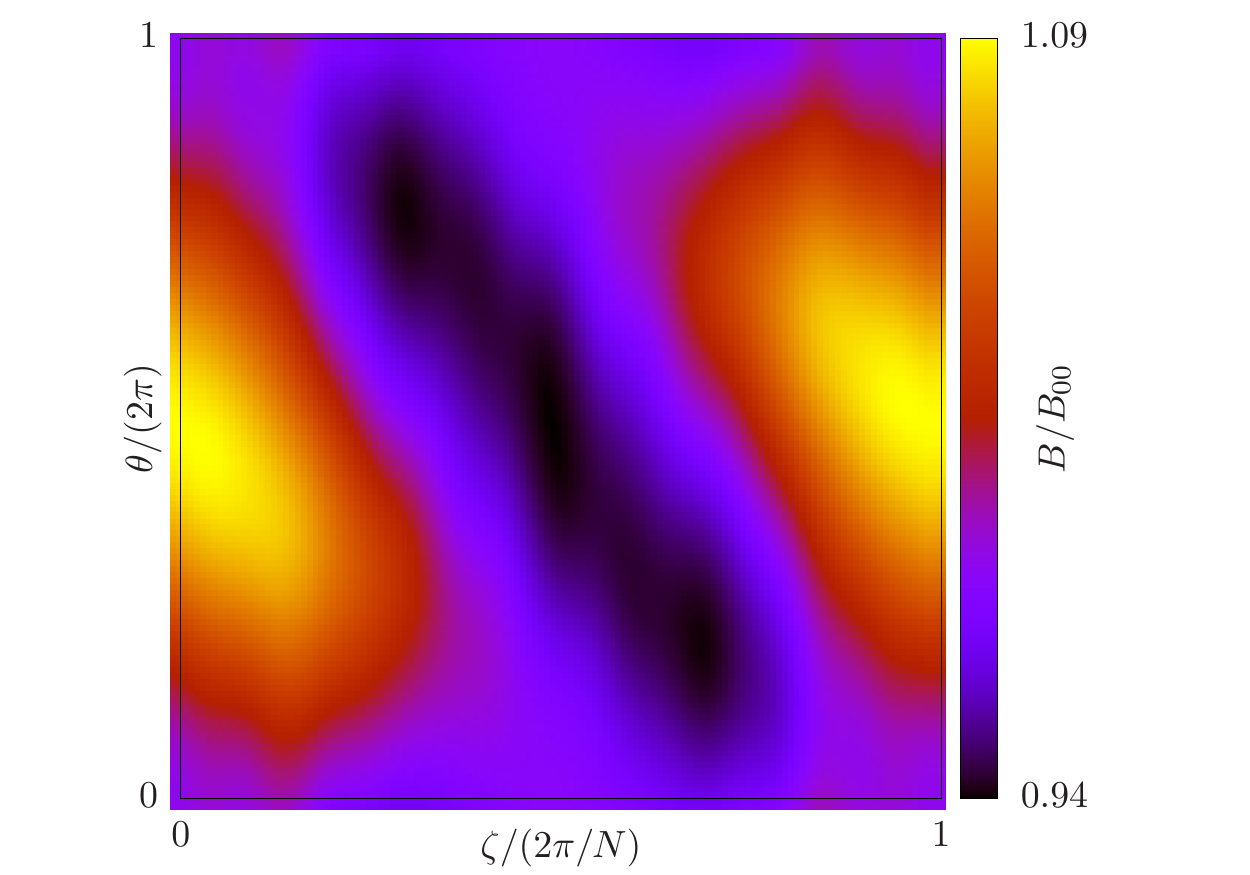}
\includegraphics[angle=0,width=0.5\columnwidth]{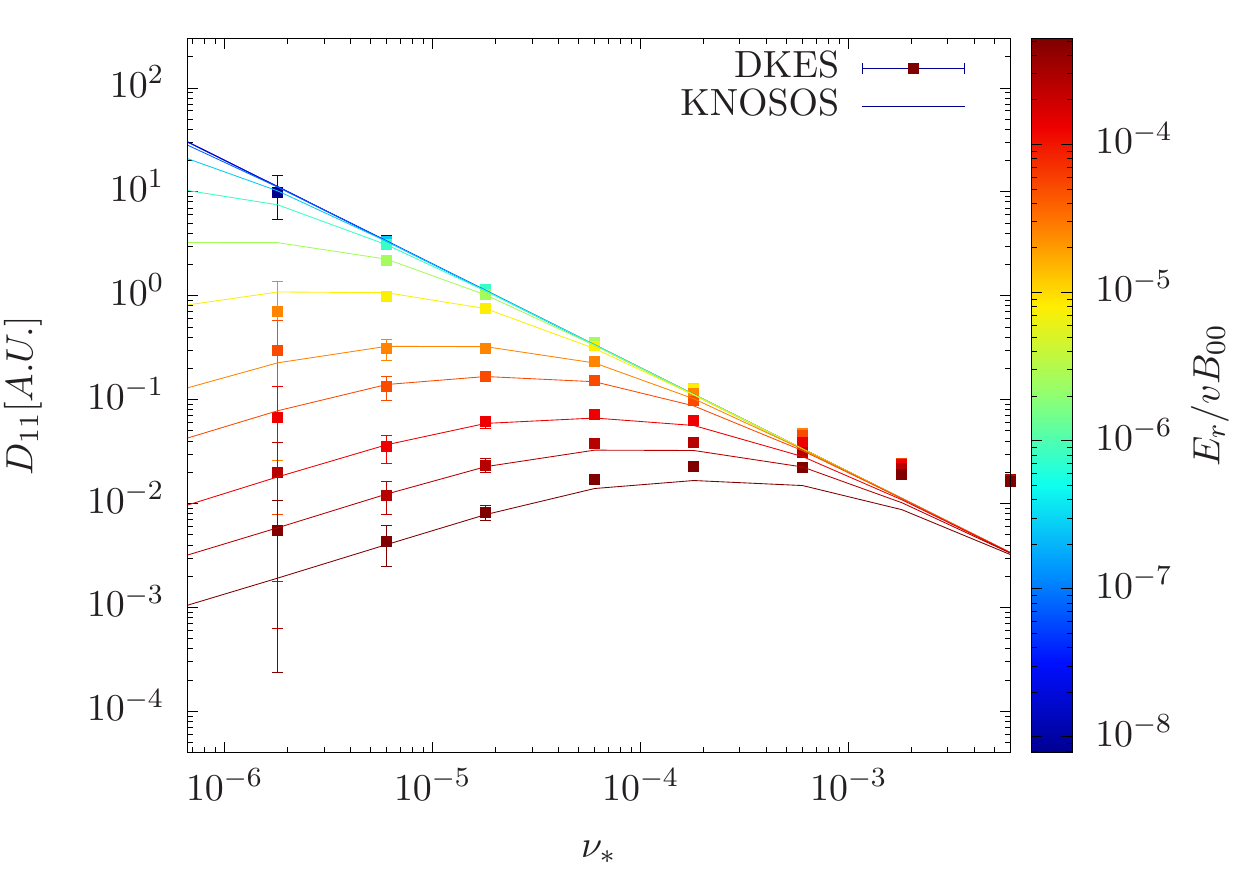}
\caption{Magnetic field strength on the flux-surface (left) and monoenergetic radial transport coefficient $D_{11}$ as a function of the collisionality and normalized radial electric field (right). The calculations correspond to configurations FTM (top) and DBM (bottom) of Wendelstein 7-X.}
\label{FIG_D11}
\end{figure}

\section{Results}\label{SEC_RES}

In this section, we demonstrate the performance of the new version of \texttt{KNOSOS} . In~\ref{SEC_DKES} we benchmark it against \texttt{DKES} in the large aspect ratio limit. In section~\ref{SEC_SBP}, we show the effect of the component of the magnetic drift that is tangent to magnetic surfaces in the analysis of stellarator transport.

\subsection{Fast and accurate calculation at low $\nu_*$  and $E_r$ of standard size}\label{SEC_DKES} 

In this section, we deal with the neoclassical transport of bulk species $b$ in large aspect ratio stellarators ($\epsilon\ll 1$) for radial electric fields of size $E_r \sim T_b/(a\,Z_be)$ (this is the standard size of a radial electric field that is solution of an ambipolarity equation in which bulk species $b$ intervenes). As a consequence, in equation~(\ref{EQ_DKE}), the terms including $\mathbf{v}_{M,b}\cdot\nabla \alpha$ and $\mathbf{v}_E\cdot\nabla \psi$ are negligible. We are left with

\

\begin{eqnarray}
\int_{l_{b_1}}^{l_{b_2}} \frac{\mathrm{d}l}{|v_\parallel|} \frac{B}{\fsa{B}}\mathbf{v}_E\cdot\nabla\alpha \left(\partial_\alpha + \partial_\alpha\lambda |_J\partial_\lambda\right) g_b- \int_{l_{b_1}}^{l_{b_2}} \frac{\mathrm{d}l}{|v_\parallel|} C_b^{\mathrm{lin}}[g_b] =\nonumber\\~~~~~~~~~~~~~~~~~~~~~~~~~~~~~~~~~~~~~~~~~-\int_{l_{b_1}}^{l_{b_2}} \frac{\mathrm{d}l}{|v_\parallel|} \mathbf{v}_{M,b}\cdot\nabla \psi \Upsilon_b F_{M,b}\,. 
\label{EQ_BDKES}
\end{eqnarray}
This equation rigorously models the transport of large aspect-ratio stellarators at low collisionality and $E_r$ of standard size~\cite{dherbemont2021fow}. Specifically, it describes the $1/\nu$ regime (when $\epsilon^{3/2}\gg\nu_*\gg \rho_*/\epsilon$) and the $\sqrt{\nu}$ or $\nu$ regimes ($\nu_*\ll \rho_*/\epsilon$). For large aspect ratio, this equation coincides with the orbit average of the equation solved by \texttt{DKES} (including the incompressible $E\times B$ drift \cite{hirshman1986dkes,beidler2007icnts}, which is correct when $\epsilon\ll 1$ and $E_r \sim T_b/(a\,Z_be)$). More details are given in~\ref{SEC_APP}.

In order to benchmark \texttt{KNOSOS}, we perform calculations of the monoenergetic transport coefficient $D_{11}$ as a function of the collisionality and normalized radial electric field. We do so for 5 magnetic configurations (7 different flux-surfaces in each case, labelled by $s={\psi}/{\psi_{LCMS}}=0.005, 0.046, 0.128, 0.250, 0.413, 0.617$ and 0.862, where $\psi=\psi_{LCMS}$ at the last closed flux-surface) in the configuration space of the stellarator Wendelstein 7-X: low-mirror (officially labelled AIM), standard (EIM), high-mirror (KJM), high-iota (FTM) and low-iota (DBM). This set comprises configurations with varying degree of optimization with respect to neoclassical transport. Figures~\ref{FIG_D11b} and ~\ref{FIG_D11}~show a selection of the calculations, corresponding to $s=0.128$. The agreement with~\texttt{DKES} is good, and the computing time is much smaller: the full characterization of the flux-surface takes a few seconds, while it may need up to tens of hours with~\texttt{DKES}.

\begin{figure}[!ht]
\includegraphics[angle=0,width=0.5\columnwidth]{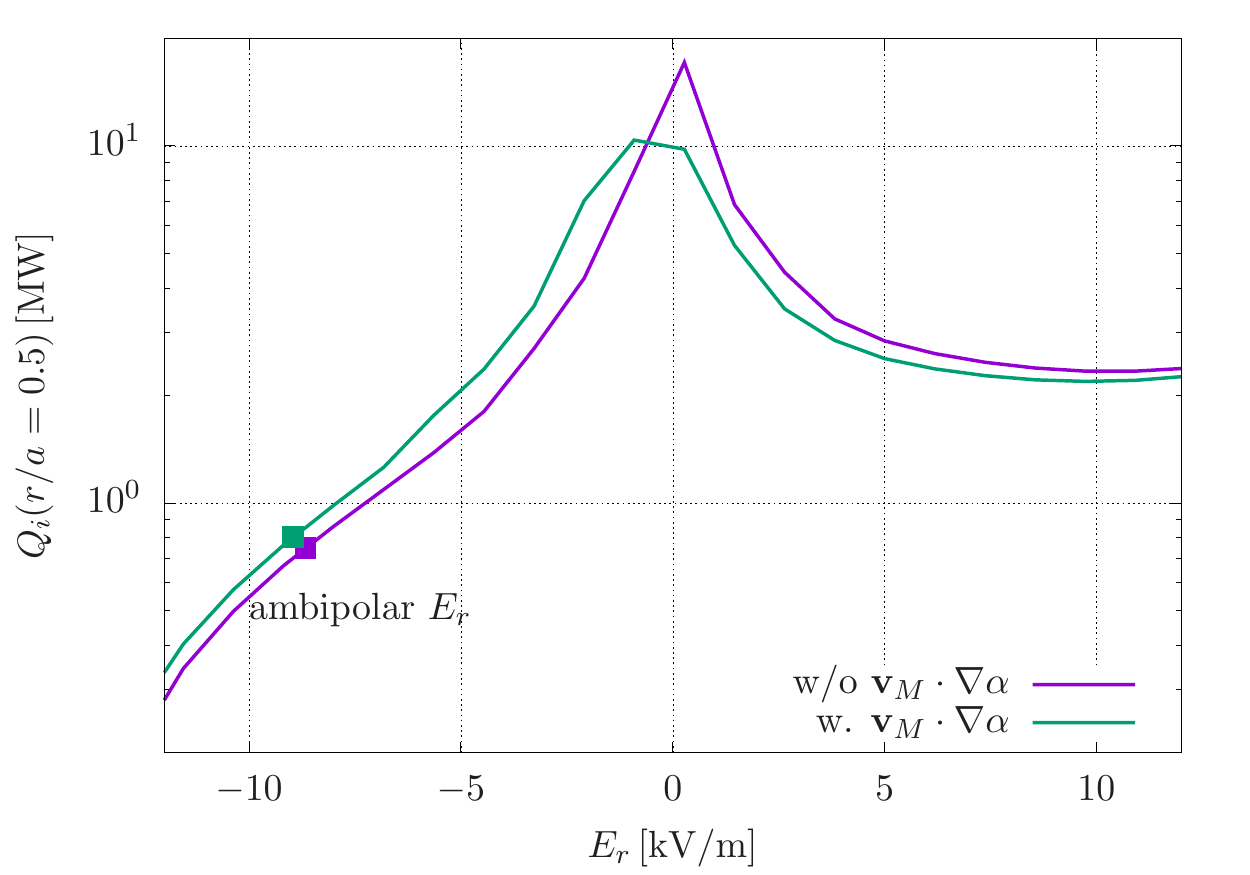}
\includegraphics[angle=0,width=0.5\columnwidth]{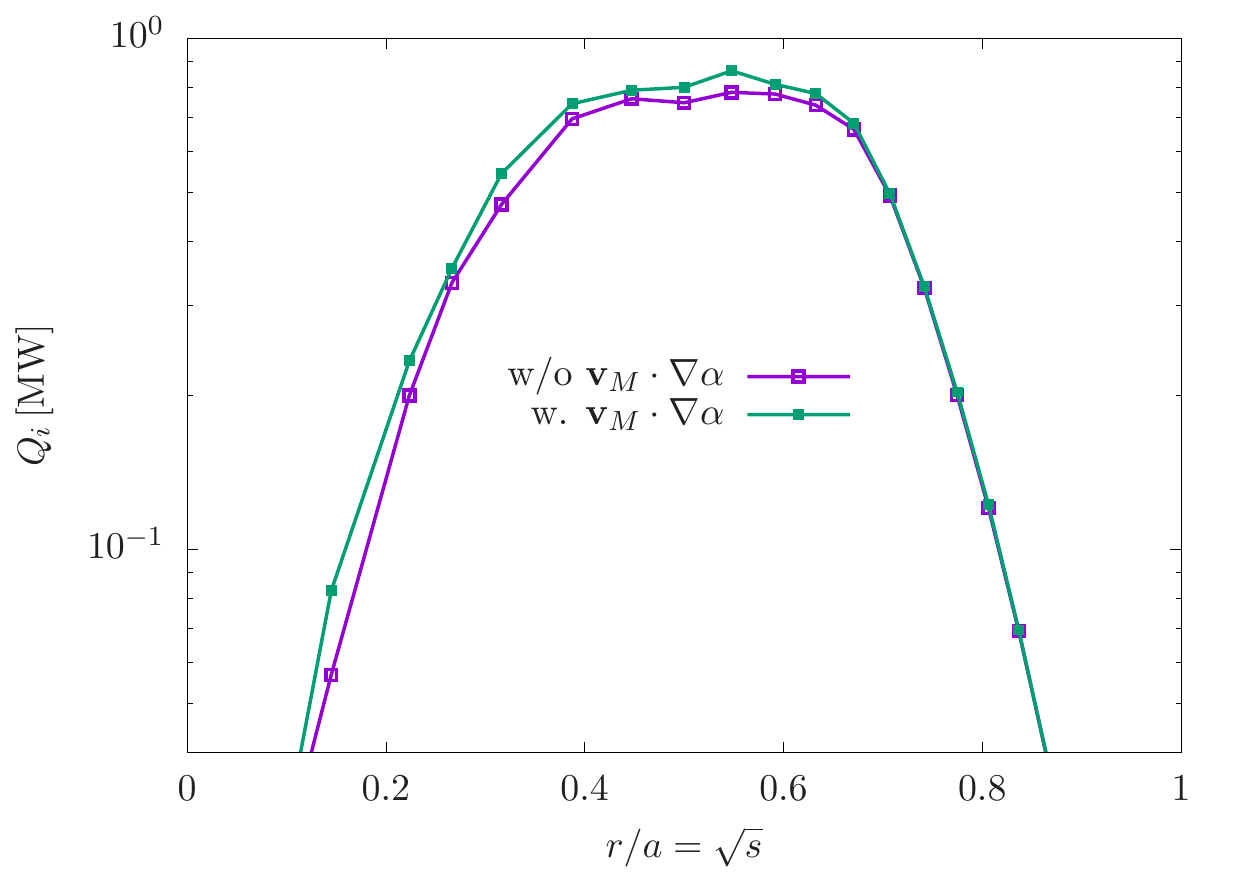}\\
\begin{center}
\vskip-0.5cm~~~~~~\includegraphics[angle=0,width=0.5\columnwidth]{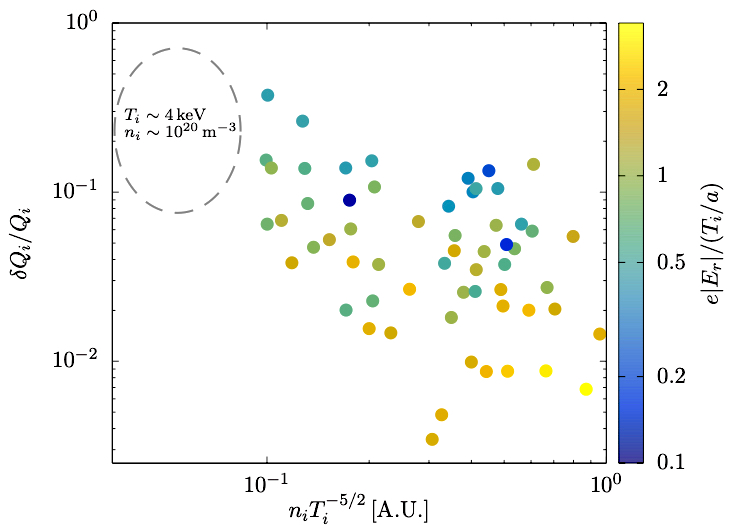}
\end{center}
\caption{Neoclassical ion energy flux as a function of the radial electric field at radial position $r/a=0.5$ (top left), and as a function of the radial position (top right). Relative change in the ion energy flux caused by the tangential magnetic drift (bottom).}
\label{FIG_VM}
\end{figure}

\subsection{Fast and accurate calculation at low $\nu_*$  and small $E_r$}\label{SEC_SBP} 

In this section, we address the neoclassical transport of large aspect ratio stellarators for small radial electric fields ($E_r \ll T_b/(a\,Z_be)$). We will assume that the stellarator is close enough to omnigeneity, so that a local equation can be employed. As a consequence, in equation~(\ref{EQ_DKE}), the terms $\mathbf{v}_{M,b}\cdot\nabla \alpha$ and $\mathbf{v}_E\cdot\nabla \psi$ are not negligible, but the term $\partial_\alpha\lambda |_J\partial_\lambda$  is~\cite{calvo2017sqrtnu}. This yields
\begin{eqnarray}
\int_{l_{b_1}}^{l_{b_2}} \frac{\mathrm{d}l}{|v_\parallel|} \left(\mathbf{v}_{M,b} + \frac{B}{\fsa{B}}\mathbf{v}_E\right)\cdot\nabla\alpha \partial_\alpha g_b - \int_{l_{b_1}}^{l_{b_2}} \frac{\mathrm{d}l}{|v_\parallel|} C_b^{\mathrm{lin}}[g_b] = \nonumber~~~~~~~~~~~~~~~~~~~~~\\
=-\int_{l_{b_1}}^{l_{b_2}} \frac{\mathrm{d}l}{|v_\parallel|} (\mathbf{v}_{M,b}+\mathbf{v}_E)\cdot\nabla \psi \Upsilon_b F_{M,b}\,,~~~~~~~~~~~~~~~~~~~~~\
\label{EQ_ODKE}
\end{eqnarray}
This equation models rigorously the transport of optimized large-aspect ratio stellarators at low collisionality \cite{velasco2020knosos}. Specifically, for small $E_r$, it describes the $1/\nu$ regime (for $\epsilon^{3/2}\gg\nu_*\gg \rho_*$) and the superbanana-plateau or $\sqrt{\nu}$ regimes (for $\nu_*\ll \rho_*$ and depending on details of the magnetic field~\cite{calvo2018jpp}); for large $E_r$, it describes the $1/\nu$ regime (when $\epsilon^{3/2}\gg\nu_*\gg \rho_*/\epsilon$) and the $\sqrt{\nu}$ regime ($\nu_*\ll \rho_*/\epsilon$). We note that even though the term with $\partial_\alpha\lambda |_J\partial_\lambda$ is small in this limit, retaining it can be useful in order to deal numerically with very deeply trapped particles~\cite{velasco2020knosos}.

In order to assess the effect of tangential magnetic drift on transport, we compare two calculations: in the first one, we solve equation~(\ref{EQ_DKE}) with $\varphi_1$ set to zero; in the second one, we solve equation (\ref{EQ_BDKES}). We perform the calculations using as input the experimental profiles and magnetic configuration (EIM) of a W7-X discharge: \#180918041, studied in detail in~\cite{estrada2021op12}. It corresponds to a so-called "high-performance" plasma, in which the ion turbulence has been reduced and neoclassics provides the main contribution to ion energy transport in the core region~\cite{bozhenkov2020hp}. Figure~\ref{FIG_VM} (top left) shows this neoclassical $Q_i$ as a function of the radial electric field at radial position $r/a=0.5$. A well-known qualitative behaviour is observed~\cite{matsuoka2015tangential}: the large peak in $Q_i$ that appears at $E_r=0$ when solving equation (\ref{EQ_BDKES}) (or, equivalently, when using~\texttt{DKES}) is reduced in height and moves towards negative values of $E_r$ (an opposite effect takes place, not shown in the figure, for the electrons). The difference $\delta Q_i$ between the two calculations for a given value of $E_r$ is thus strongest for small $E_r$. Figure~\ref{FIG_VM} (top right) shows the radial profile of neoclassical $Q_i$ for discharge: \#180918041. It can be observed that the neoclassical predictions change quantitatively, but not qualitatively, with $\delta Q_i$ around 10\% at $r/a<0.5$ (and negligible for outer positions, where the collisionality is large). The reason is that the ambipolar value of $E_r$ is never too small. Furthermore, as indicated by the squares in figure~\ref{FIG_VM} (top left), the small change in the ambipolar radial electric field goes in the direction of making $\delta Q_i$ small.

Even if the effect is small, and very likely lies within the precision of the calculations (that come e.g. from the error bars in the measured plasma profiles), it is systematic, and is likely to become larger in other relevant scenarios. In W7-X, the inverse aspect ratio is very small, and therefore very small $E_r$ is needed in order for the tangential magnetic drift to become of the size of the $E\times B$ drift. This is not the case of the Large Helical Device, see e.g.~\cite{matsuoka2015tangential,velasco2020knosos}. And even within the configuration space of W7-X, there is a variety of scenarios. For the high-mirror (KJM) configuration of W7-X, the tangential magnetic drift is relatively larger, specially at higher $\beta$ due to the diamagnetic effect (the confinement of energetic ions relies precisely on enhancing their precession on the flux-surface). Finally, for any configuration, an accurate description of the particle orbits will be necessary at high enough temperature. In order to quantify this, in figure~\ref{FIG_VM} (bottom) we repeat the calculation of figure~\ref{FIG_VM} (top left) for 8 W7-X discharges selected from~\cite{carralero2021EPS}. Two trends can be observed: $\delta Q_i/Q_i$ increases with decreasing $n_iT_i^{-5/2}$ (i.e. for decreasing weight of collisions with respect to the drifts on the flux-surface) and decreasing $e|E_r|/(aT_i)$ (a value $\sim 1$ indicates a radial electric field of standard size). If forthcoming campaigns of W7-X are able to reach ion temperatures of the order of 4$\,$keV, $\delta Q_i/Q_i\sim 1$ is a possibility to be considered.

\section{Summary}\label{SEC_SUM}

In this work, we have presented a new version of \texttt{KNOSOS}, a powerful low-collisionality neoclassical code that is extremely fast, and at the same time, can calculate physical effects usually neglected: tangential magnetic drift and $\varphi_1$. Thanks to the implementation of new bounce-averaged equations, \texttt{KNOSOS} can now handle any large aspect ratio stellarator magnetic configuration. There is a wide variety of plasma physics problems that it can be applied to. A non exhaustive list includes the analysis of experimental discharges, stellarator optimization and the provision of input ($E_r$, $\varphi_1$ or the neoclassical distribution function $g_b$) for other transport simulations.

\section*{Acknowledgments}

The work has been partially funded by the Ministerio de Ciencia, Innovaci\'on y Universidades of Spain under project PGC2018-095307-B-I00 and FIS2017-88892-P and by the Comunidad de Madrid under projects 2017-T1/AMB-5625 and Y2018/NMT [PROMETEO-CM]. This work has been carried out within the framework of the EUROfusion Consortium and has received funding from the Euratom research and training programme 2014-2018 and 2019-2020 under grant agreement No 633053. The views and opinions expressed herein do not necessarily reflect those of the European Commission.

\appendix

\section{Discussion on the drift-kinetic equation for large aspect ratio stellarators}\label{SEC_APP}

In this appendix we discuss how equation (\ref{EQ_BDKES}) coincides at low collisionalities and large aspect ratio with the drift kinetic-equation solved by the code~\texttt{DKES}. Let $h(\psi,\alpha,l,v,\lambda,\sigma)$ be the non-adiabatic component of the deviation of the ion distribution from a Maxwellian distribution (in this appendix we drop the species index). The function $h$ is defined in the trapped and passing regions of the phase space; it depends on $l$ (it is not an orbit-averaged quantity) and $\sigma$ (i.e., it is not even in the parallel velocity). \texttt{DKES} computes $h$ by solving~\cite{hirshman1986dkes}
\begin{equation}
\frac{B^2}{\fsa{B^2}}\mathbf{v}_E\cdot\left(\nabla h - \frac{\lambda}{B}\nabla B\partial_\lambda h\right) + v_\parallel \frac{\mathbf{B}}{B}\cdot\nabla h + \mathbf{v}_{M}\cdot\nabla \psi \Upsilon F_{M} = C^{\mathrm{lin}}[h]\,.~~~~~
\label{EQ_DKES}
\end{equation}
Equation (\ref{EQ_DKES})~is typically solved imposing the following regularity conditions in the derivatives with respect to $\lambda$:
\begin{equation}
\lambda\partial_\lambda h=0~\mathrm{at}~\lambda=0
\label{EQ_B1}
\end{equation}
and
\begin{equation}
|v_\parallel\partial_\lambda h| < \infty~\mathrm{at~points~with}~v_\parallel=0\,.
\label{EQ_B2}
\end{equation}
Note that equation (\ref{EQ_DKES}), together with the boundary conditions (\ref{EQ_B1}) and (\ref{EQ_B2}), do not completely determine $h$: one can add an arbitrary function of $\psi$ and $v$ to $h$ and obtain another solution of (\ref{EQ_DKE}) that also satisfies (\ref{EQ_B1}) and (\ref{EQ_B2}). This function is typically unimportant (it does not give radial transport, for example).

In equation~(\ref{EQ_DKES}),  $(B^2/\fsa{B^2})\mathbf{v}_E$ is an incompressible approximation to the $E\times B$ drift $\mathbf{v}_E$, and this allows \texttt{DKES} to solve the drift-kinetic equation by means of a variational principle. This approximation is correct for $\epsilon\ll 1$, except for very large values of radial electric field, $E_r\gg T/(a\,Ze)$~\cite{beidler2007icnts}. For $\epsilon\ll 1$, equation (\ref{EQ_DKES}) is equivalent to
\begin{equation}
\mathbf{v}_E\cdot\left(\nabla h - \frac{\lambda}{\fsa{B}}\nabla B\partial_\lambda h\right) + v_\parallel \frac{\mathbf{B}}{B}\cdot\nabla h + \mathbf{v}_{M,b}\cdot\nabla \psi \Upsilon F_{M,b} = C^{\mathrm{lin}}[h]\,.~~~~~~~~~~~
\label{EQ_DKEC}
\end{equation}

The orbit averaging of equation (\ref{EQ_DKEC}) yields
 \begin{eqnarray}
~~\overline{\mathbf{v}_E\cdot\nabla l\left(\partial_l h - \frac{\lambda}{\fsa{B}} \partial_\alpha B\partial_\lambda h\right)} + \nonumber \\ 
+\mathbf{v}_E\cdot\nabla\alpha\left(\overline{\partial_\alpha h} -  \frac{\lambda}{\fsa{B}} \overline{\partial_\alpha B\partial_\lambda h}\right) +\overline{\mathbf{v}_{M}\cdot\nabla \psi} \Upsilon F_{M} = \overline{C^{\mathrm{lin}}[h]}\,,
\end{eqnarray}
where we have used that $\mathbf{v}_E\cdot\nabla\alpha=\varphi_0'/\Psi_t'$ is constant on the flux-surface, and the orbit average of a function $f$ is defined by equation~(\ref{EQ_BAV}) for trapped particles and by
\begin{eqnarray}
\overline{f}=\frac{\fsa{Bf/v_\parallel}}{\fsa{B/v_\parallel}}\,.
\label{EQ_BINTP}
\end{eqnarray}
for passing particles. The function $h$ can be decomposed into its bounce-average and a component that fluctuates along the orbit
\begin{equation}
h=\overline{h}  + h^f\,,
\end{equation}
which leads to
 \begin{eqnarray}
\overline{(\mathbf{v}_E\cdot\nabla l)^f\partial_l {h^f}}- \frac{\lambda}{\fsa{B}}\overline{\partial_l (\mathbf{v}_E\cdot\nabla l)B}\partial_\lambda \overline{h}
- \frac{\lambda}{\fsa{B}}\overline{(\partial_l (\mathbf{v}_E\cdot\nabla l)B)^f\partial_\lambda  h^f}\nonumber\\
+\mathbf{v}_E\cdot\nabla\alpha\left(\partial_\alpha\overline{h} -  \frac{\lambda}{\fsa{B}} \overline{\partial_\alpha B}\partial_\lambda \overline{h}- \frac{\lambda}{\fsa{B}} \overline{(\partial_\alpha B)^f\partial_\lambda h^f}\right) \nonumber\\+\overline{\mathbf{v}_{M}\cdot\nabla \psi} \Upsilon F_{M} = \overline{C^{\mathrm{lin}}[\overline{h}+h^f]}\,,
\label{EQ_BDKESF}
\end{eqnarray}
where we have used $\partial_l \overline{h}=0$. For large aspect ratio stellarators, it is shown in~\cite{dherbemont2021fow} that the contribution of passing particles to radial transport is small at low collisionality, so equation~(\ref{EQ_BDKESF}) needs to be solved in the trapped region only. Furthermore, several terms in~(\ref{EQ_BDKESF}) are demonstrated to be small in  $\epsilon\ll 1$, which leads to
 \begin{eqnarray}
\mathbf{v}_E\cdot\nabla\alpha\left(\partial_\alpha\overline{h} - \frac{\lambda}{\fsa{B}} \overline{\partial_\alpha B}\partial_\lambda \overline{h}\right)+\overline{\mathbf{v}_{M}\cdot\nabla \psi} \Upsilon F_{M} = \overline{C^{\mathrm{lin}}[{h}]}\,.
\label{EQ_BDKESA}
\end{eqnarray}
We have thus arrived at equation~(\ref{EQ_BDKESA}), which differs from equation (\ref{EQ_BDKES}) (with $\overline{h}=g_b$) by terms that are small in $\epsilon\ll 1$. Even though equation~(\ref{EQ_BDKESA}) is obtained more naturally from equation~(\ref{EQ_DKEC}), equation (\ref{EQ_BDKES}) posseses a desirable property: it conserves the second adiabatic invariant $J$ exactly, as discussed in section~\ref{SEC_EQ}, and not only up to small terms in $\epsilon$.


\

\

\bibliographystyle{unsrt}
\bibliography{velasco_etal_IAEA2021.bbl}

\end{document}